\documentclass{aa}
\usepackage{txfonts}
\usepackage{graphicx}
\usepackage[authoryear,longnamesfirst]{natbib}
\usepackage{longtable}
\begin{document}

\title{A multi-wavelength census of star formation activity in the 
young embedded cluster around Serpens/G3-G6 
\thanks{Based on observations with ISO, an ESA project with 
instruments funded by ESA Member States (especially the PI 
countries: France, Germany, the Netherlands, and the United Kingdom) 
and with the participation of ISAS and NASA.}
\thanks{Tables~\ref{tab1}, \ref{tab2}, and \ref{tab3} are only 
available in electronic form at the CDS via anonymous ftp to 
cdsarc.u-strasbg.fr (130.79.128.5) or via 
http://cdsweb.u-strasbg.fr/cgi-bin/qcat?J/A+A/}
}

\author{A.A. Djupvik\inst{1} \and
        Ph. Andr\'e\inst{2,5}   \and 
        S. Bontemps\inst{3} \and 
        F. Motte\inst{2,5} \and
        G. Olofsson\inst{4} \and 
        M. G\aa lfalk\inst{4} \and
        H.-G. Flor\'en\inst{4}}

\institute{ Nordic Optical Telescope, Apdo 474, 38700 Santa Cruz de 
            La Palma, Spain
  \and CEA/DSM/DAPNIA, Service d'Astrophysique, C.E. Saclay, Orme des 
       Merisiers, 91191 Gif-sur-Yvette, France
  \and Observatoire de Bordeaux, BP 89, 33270 Floirac, France
  \and Stockholm Observatory, Roslagstullsbacken 21, 10691 Stockholm, 
       Sweden
  \and AIM -- Unit\'e Mixte de Recherche CEA -- CNRS -- Universit\'e 
       Paris VII -- UMR 7158, France
       }

\offprints{A.A. Djupvik (before: A.A. Kaas)}
\mail{amanda@not.iac.es}

\date{Received 3 May 2006 / Accepted 6 July 2006}

\titlerunning{The young embedded cluster around Serpens/G3-G6}
\authorrunning{Djupvik et al.}

\abstract
{}
{The aim of this paper is to characterise the star formation activity 
in the poorly studied embedded cluster Serpens/G3-G6, located 
$\sim$~45$\arcmin$ (3~pc) to the south of the Serpens Cloud Core, 
and to determine the luminosity and mass functions of its population 
of Young Stellar Objects (YSOs). 
}
{Multi-wavelength broadband photometry was obtained to sample the 
near and mid-IR spectral energy distributions to separate YSOs from 
field stars and classify the YSO evolutionary stage. ISOCAM 
mapping in the two filters LW2 (5-8.5 $\mu$m) and LW3 (12-18 $\mu$m) 
of a $19\arcmin \times 16\arcmin$ field was combined with JHK$_S$ 
data from 2MASS, K$_S$ data from Arnica/NOT, and L$\arcmin$ data from 
SIRCA/NOT. Continuum emission at 1.3~mm (IRAM) and 3.6~cm (VLA) was 
mapped to study the cloud structure and the coldest/youngest sources.
Deep narrow band imaging at the 2.12~$\mu$m S(1) line of H$_2$ from 
NOTCam/NOT was obtained to search for signs of bipolar outflows.
}
{We have strong evidence for a stellar population of 31 Class\,II 
sources, 5 flat-spectrum sources, 5 Class\,I sources, and two 
Class\,0 sources. Our method does not sample the Class\,III 
sources. The cloud is composed of two main dense clumps aligned 
along a ridge over $\sim$~0.5~pc plus a starless core coinciding 
with absorption features seen in the ISOCAM maps. We find two
S-shaped bipolar collimated flows embedded in the NE clump,
and propose the two driving sources to be a Class\,0 candidate 
(MMS3) and a double Class\,I (MMS2). For the Class\,II population
we find a best age of $\sim$~2~Myr and compatibility with recent
Initial Mass Functions (IMFs) by comparing the observed Class\,II 
luminosity function (LF), which is complete to 0.08~$L_{\odot}$, 
to various model LFs with different star formation scenarios and
input IMFs. 
}
{} 


\maketitle


\section{Introduction}
\label{intro}

The Serpens Cauda Clouds are among the darkest regions in a complex 
of clouds called the Aquila Rift \citep{dam85}. The young embedded 
cluster in the Serpens Cloud Core has been well studied at most 
wavelengths, see \citet{eir06} for a recent review. About 45' to the 
south there is also a region with active star formation. \citet{coh79} 
found 4 optical T Tauri stars in a 30$\arcsec$ field and named them 
Ser/G3-~G6. \citet{cla90,cla91} mapped the area surrounding Ser/G3-~G6 
in the ammonia 1,1 emission line and found two NH$_3$ cores, one on 
each side of the optically visible stellar group: Ser/G3-~6NE and 
Ser/G3-~G6SW. \citet{zie99} surveyed 3.14 square degrees in the 
optical lines of [S\,II] and found the Herbig-Haro object HH\,476 
close to the Ser/G3-G6SW core. The energy source of HH\,476 was 
identified by \citet{wu02} to be IRAS18265+0028. At this position 
\citet{per94} found a H$_{2}$O maser. Recently, \citet{har06} found
24 YSO candidates in a 12' $\times$ 12' area in this region 
(which they refer to as Cluster B) from a mid-IR survey with Spitzer.

When preparing the {\it ISOCAM star formation survey},\footnote{One 
of the guaranteed time programs with the camera aboard the Infrared 
Space Observatory (ISO).} this region was defined based on the 
ammonia maps of \citet{cla90} and named {\em Serpens NH3}. The field 
is relatively opaque at optical wavelengths and has a peak brightness 
at 60 $\mu$m \citep{zha88}.

We will use the distance of 225 $\pm$ 55 pc to the Serpens Cauda 
Clouds according to \citet{str03}, keeping in mind the caveat of 
the same authors that the cloud complex is possibly 80 pc deep.

Preliminary ISOCAM results from this region were presented in 
\citet{kaa99c} and \citet{kaa99b}.
In this paper we present the full results of a 17'$\times$ 19' ISOCAM 
survey in LW2 (6.7 $\mu$m) and LW3 (14.3 $\mu$m), cross-correlated 
with the 2MASS point source catalogue. In addition we present deep 
follow-up $H_2$ line (2.122 $\mu$m), $K_S$ band (2.15 $\mu$m), and 
$L'$ band (3.8 $\mu$m) imaging from the Nordic Optical Telescope, 
as well as IRAM 1.3~mm continuum and VLA 3.6~cm mapping of the most 
active part of the region.

We describe the various observation sets with reductions and 
results in Sect.~\ref{obs}, the use of IR excesses to extract 
YSO candidates in Sect.~\ref{irex}, the classification of the 
YSO population in Sect.~\ref{ysopop}, the luminosity and mass 
function in Sect.~\ref{lfmf}, and the cloud structure and the 
outflow sources in Sect.~\ref{flow}.

%
\section{Observations and reductions}
\label{obs}

\subsection{ISOCAM}
\label{obsiso}

This paper is based on observations obtained with ISOCAM, the camera
aboard the Infrared Space Observatory (ISO; see 
\citealt{kes96,ces96}), 
and is part of the LNORDH.SURVEY\_1 star formation survey conducted 
in the two broadband filters LW2 (5-8.5 $\mu$m) and LW3(12-18 $\mu$m), 
see, e.g., \citet{kaa00} for a review. 
In the region called Serp-NH3 about 0.09 square degrees were 
surveyed in LW2 and LW3 in 1996. The pixel field of 
view (PFOV) was 6$\arcsec$ and the unit integration time 0.28 
seconds. The region was mapped by rastering along right ascension 
with about half a frame (90$\arcsec$) overlap in $\alpha$ and 
24$\arcsec$ overlap in $\delta$, producing a 12 by 6 mosaic. Each 
position in the sky was observed for about 15 seconds. Image 
reductions were performed using 
CIA\footnote{CAM Interactive Analysis, a joint development by the ESA 
Astrophysics Division and the ISOCAM Consortium led by the ISOCAM PI,
C. Cesarsky, Direction des Sciences de la Matiere, C.E.A., France.} 
and a set of our own programs for point source detection and 
photometry. 
We refer to \citet{kaa04} for a detailed description of the data 
reduction. The fluxes in ADU/s in LW2 and LW3 are converted to mJy 
through the relations 2.32 and 1.96 ADU/gain/s/mJy for LW2 and LW3, 
respectively \citep{blo00}. These are strictly valid only for 
F$_{\nu} \propto \nu^{-1}$ sources and a small colour correction 
has been applied to the ``blue'' sources, which have significantly
different spectra (F$_{\nu} \propto \nu^{-3}$).
The reference wavelengths are defined at 6.7 and 14.3 $\mu$m for LW2 
and LW3, respectively. Conversion from flux density to magnitude is 
defined as m$_{\rm 6.7} = -2.5 \log (F_{\nu}(6.7 \mu m)/82.8)$ and 
m$_{\rm 14.3 \mu m} = -2.5 \log (F_{\nu}(14.3 \mu m)/18.9)$, where 
$F_{\nu}$ 
is given in Jy. The $\sim 5$ \% responsivity decrease throughout 
orbit has not been corrected for. The ISOCAM astrometry was based on 
the ISO pointing only. We manually cross-correlated bright sources 
with 2MASS positions and found a bulk offset of 7'' in DEC and 1.4'' 
in RA. Correcting for this, the mean deviation between ISOCAM and 
2MASS positions (for these bright sources) became 2.5 $\pm$ 1.4''. 

ISOCAM point source extraction and photometry gave a total of 186 
detections (see Table~\ref{tab1}) of which 160 have 
reliable\footnote{Detections without reliable flux measurements are 
generally sources observed on the edge of the detector.} fluxes in 
the 6.7 $\mu$m band (LW2) and 74 in the 14.3 $\mu$m band (LW3). 
For the 14.3 $\mu$m band we estimate a completeness limit of 8 mJy 
and a limiting sensitivity of 3 mJy, while for the 6.7 $\mu$m band 
the survey is complete to 6 mJy and reaches 1 mJy. At these levels
the sample is not contaminated by galaxies. See the discussion in 
\citet{kaa04} for details.

\begin{figure*}
\resizebox{\hsize}{!}{\includegraphics{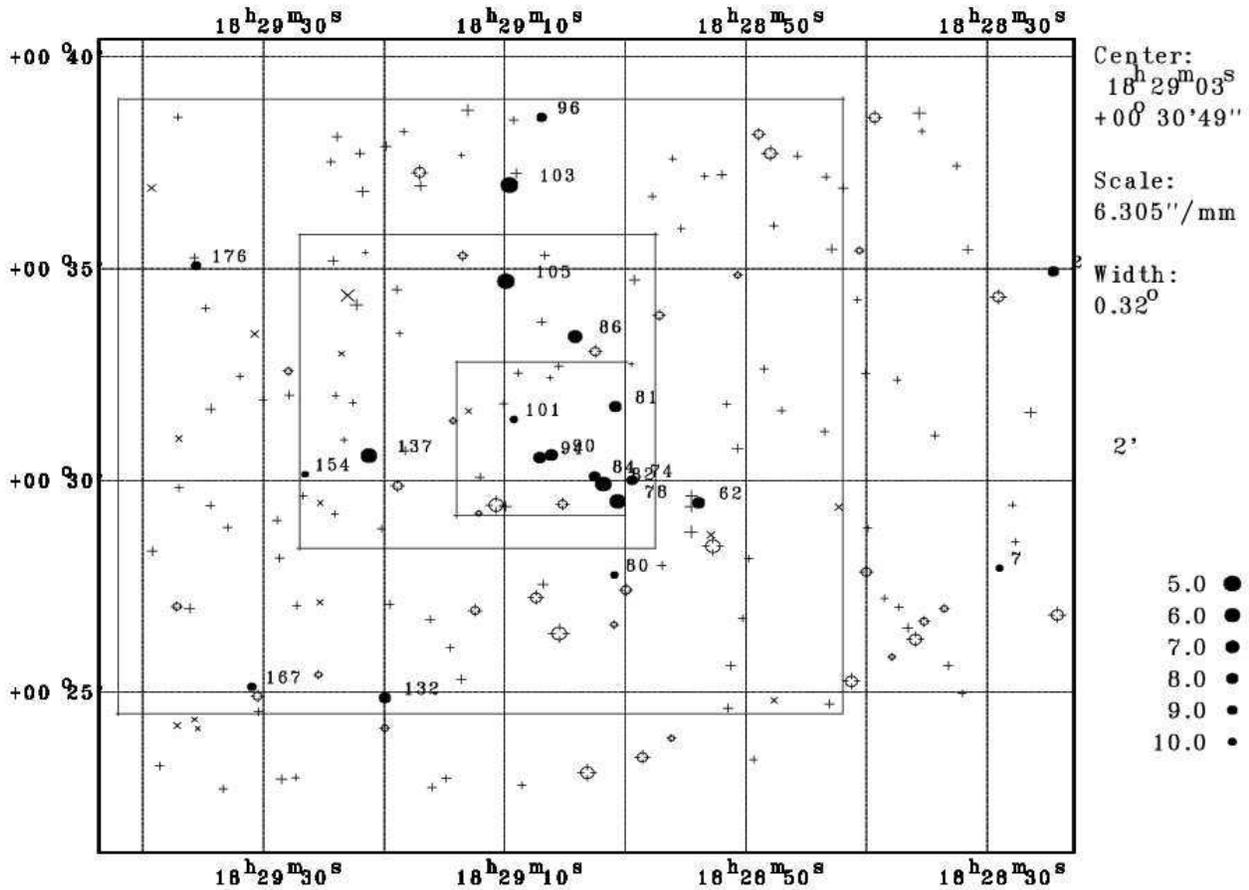}}
\caption{Sky map of the ISOCAM sources. Symbol size inversely
proportional to LW2 magnitudes. IR-excess sources (filled circles), 
sources without IR-excess or ``blue'' sources (open circles), only
LW2 fluxes (plus sign), and only LW3 fluxes (crosses). Coordinates
are J2000 epoch. The three subregions indicated are the areas 
observed with IRAM (large box), Arnica (medium box), and NOTCam
(small box). The pointed SIRCA observations are not shown. See text 
for details.}
\label{fig-isomap}
\end{figure*}

Table~\ref{tab1}, which is available in electronic form at the CDS, 
lists all the 186 ISOCAM detections with fluxes 
and uncertainties for each band (see Sect.~\ref{irex} for a 
description of the columns).
Flux measurements have been obtained in both bands for a total of
57 sources. These belong to either of two colour populations, one 
group of 36 ``blue'' sources that have colour indices like normal 
photospheres, and another group of 21 ``red'' sources that have 
excess emission at 14.3 $\mu$m (see Fig.~\ref{fig-323} and 
Sect.~\ref{irex} in general for details). 
Figure~\ref{fig-isomap} is a sky map of the ISOCAM sources. The 
locations of the IR-excess sources, that are also labelled, are
shown together with those of the ``blue'' (no IR-excess) sources, 
as well as sources with either LW2 or LW3 fluxes only.
The subregions marked in this map show the location and extent of
various follow-up observations: IRAM 1.3mm map (large box), Arnica
deep Ks imaging (medium box), and NOTCam H$_2$ line imaging (small
box). The SIRCA pointed observations are not indicated in the figure.

%
\begin{table}
\caption{The 186 ISOCAM detections in the Ser/G3-G6 complex. Only
available in electronic form at the CDS.
\label{tab1}
        }
%
%
%
%
%
%
%
%
\end{table}

\subsection{2MASS}
\label{obs2m}

In the region mapped by ISOCAM a total of 1803 sources was found in 
the 2MASS All-Sky Release Point Source Catalog \citep{cut03}. The 
$J$ (1.24 $\mu$m), $H$ (1.66 $\mu$m), and $K_s$ (2.16 $\mu$m) 
photometry of all these has been used in this paper. 

\begin{figure*}
\resizebox{\hsize}{!}{\includegraphics{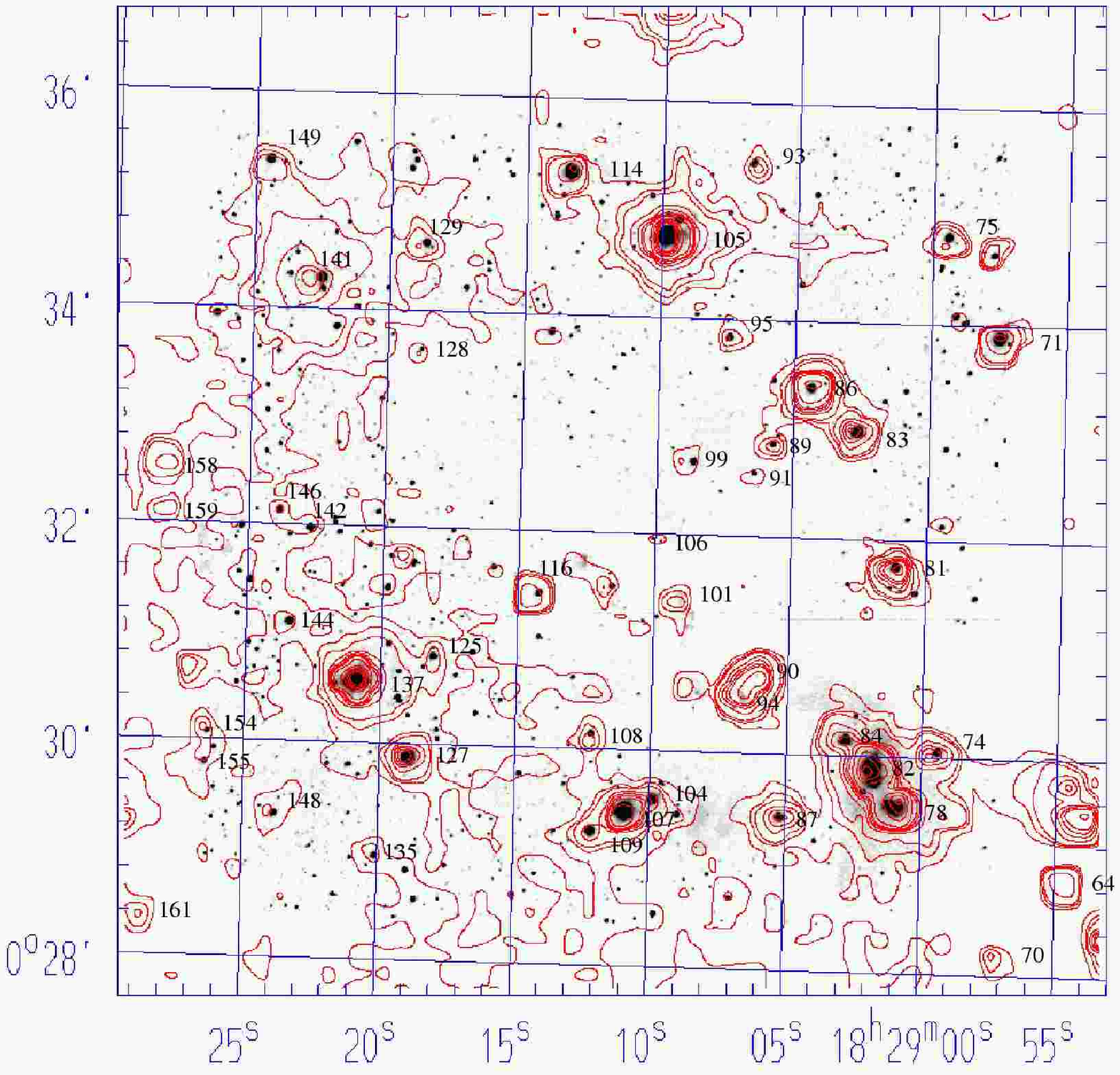}}
\caption{Arnica/NOT $K_s$ band 4 $\times$ 4 mosaic of an 8 $\times$ 
8 arc minutes part of the ISOCAM area. Contours show the ISOCAM LW2 
(6.7 $\mu$m) mosaic and ISO ID numbers are labelled (to the right
of the source).}
\label{fig-nh3-ks}
\end{figure*}

We cross-correlated the 186 ISOCAM sources with the 2MASS PSC using 
a search radius of 7.5'' (i.e., 3 times the mean positional deviation 
between ISOCAM and 2MASS for bright sources after the global
recentering, see Sect.~\ref{obsiso}).
Only 19 ISOCAM detections had no 2MASS counterpart, and of these, 
three 
are very red and faint in the near-IR, while the remaining had been 
detected in only one of the two ISOCAM bands. In 29 cases we found 
multiple 2MASS objects within the search radius. Disregarding these, 
the mean positional deviation of all cross-correlations is 3.1 $\pm$ 
1.5'' and the maximum is 6.5'', and this is a measure of the accuracy 
of the ISOCAM positions. None of the 29 ISOCAM sources that are
associated with multiple 2MASS sources within a 7.5'' radius are
YSO candidates with excess emission at 14.3 $\mu$m, although ISO-132
might be a Class\,II to Class\,III transition object (cf. 
Sect.~\ref{ysopop}). In the 28 cases of two 2MASS sources within the 
search radius, we have discarded one of them as ``probably not an 
ISOCAM counterpart'', based on separation, color, and quality flags. 
In one case of a quadruple (ISO-NH3-118), we have discarded 3 objects.

\subsection{$K_s$ band imaging with Arnica at the NOT}
\label{obsarn}

In August 1996, about 4 months after the ISOCAM observations, the 
Arcetri Near-Infrared Camera ({\sc Arnica}) was used at the Nordic 
Optical Telescope (NOT), La Palma, to obtain deep $K_s$ band (2.15 
$\mu$m) imaging in the most active $8\arcmin \times 8\arcmin$ 
subregion 
of the ISOCAM field. A 4 by 4 raster mode (256 $\times$ 256 array and 
0.54''/pix) was scanned repeatedly (dithering each scan), staring at 
the same position only 1 minute at a time with a unit integration 
time of 1 second, until a total on-source integration time of 5 
minutes was obtained for the whole map. In this way the program 
images themselves were used to 
subtract the thermal sky level. Flat fields were calibrated on the 
decreasing/increasing twilight sky, removing thermal and stray light 
contributions by taking difference images. Standard stars taken from 
the UKIRT Faint standard list (Casali \& Hawarden 1992) and from the 
{\sc Arnica} list (Hunt et al. 1998) were observed throughout the 
night. Observation and reduction methods are the same as those 
described in \citet{kaa99a}.

Figure~\ref{fig-nh3-ks} shows the deep $K_S$ band imaging with Arnica 
of an 8 $\times$ 8 arc minutes sub-region (i.e., 18 \%) of the ISOCAM 
area, and contours of the LW2 band (6.7 $\mu$m) image are overlaid. 
This area covers the Ser/G3-G6NE ammonia core \citep{cla91} well, 
and a high extinction region is outlined by the drop in the surface 
density of stars in the $K_S$ band.

Point sources were detected with daofind (threshold at 5 $\sigma$), 
and aperture photometry was made using a small aperture with an 
aperture correction applied to compensate for flux losses in the 
wings of the stellar profile. A total of 1014 sources (out of 1148 
detections) have photometry with $\sigma_{Ks} < 0.3 $ mag in the 
8' $\times$ 8' Arnica map. All are listed in Table~\ref{tab2} 
(available online only). The first column
gives the Arnica/NOT identification number from 1 to 1148. The second
and third columns list the J2000 equatorial coordinates. Columns four
and five list the $K_S$ magnitude and its error, and the last column
gives possible other identifications where cross-correlation has been
made with the ISOCAM map. The coordinates
were found by registering the mosaic using IRAF tasks {\em ccmap} and 
{\em xy2rd} with the 2MASS coordinates of 44 cross-correlated ISO 
and SIRCA sources as an input grid to the plate solution. The rms of
the J2000 coordinates presented in Table~\ref{tab2} are 0.38'' in ra
and 0.31'' in dec. This relatively large rms is probably because of 
the optical distortion in the camera, which was not corrected for. The 
sources with $\sigma_{Ks} < 0.3 $ mag are in the range from 
$K_S = 7.5$ to $K_S = 17.8$ mag, and the survey is estimated to be 
complete to $K_S \approx 16$ mag. 

%
\begin{table}
\caption{The 1148 Arnica $K_S$ band sources detected in a 8' $\times$ 
         8' subregion of ISOCAM coverage of the Ser/G3-G6 complex. 
         Only available in electronic form at the CDS.
\label{tab2}
         }
\end{table}

Comparing the $K_S$ band magnitudes of Arnica/NOT (August 1996) with 
those of 2MASS (July 2000) for 37 ISOCAM sources in the range 
$ 7.8 < K_S < 14.3 $ shows overall agreement. The difference 
$K_S (1996) - K_S (2000)$ has a median value of  -0.005 $\pm$ 0.07 
mag when disregarding six sources that have varied by more than 0.2 
mag and are therefore variable sources. These are: ISO-NH3-84 (a 
double Class\,II), ISO-NH3-86 
(flat-spectrum source), ISO-NH3-94 (Class\,I or younger - the K band 
emission is probably dominated by the emission in the 2.122 $\mu$m 
S(1) line of $H_2$), ISO-NH3-105 (Class\,II), ISO-NH3-137 (Class\,II), 
and ISO-NH3-154 (Class\,II). Refer to Tables~\ref{tab5} and 
\ref{tab6}. Because the Arnica data was collected only 4 months after 
the ISOCAM observations, we use these to calculate the $\alpha_{2-7}$ 
and $\alpha_{2-14}$ spectral indices. Also, the Arnica dataset 
resolves ISO-NH3-84 into two sources and provides a $K_S$ band
magnitude for the faint source ISO-NH3-90.

\subsection{$L'$ band imaging with SIRCA at the NOT}
\label{obssir}

\begin{figure}
\resizebox{\hsize}{!}{\includegraphics{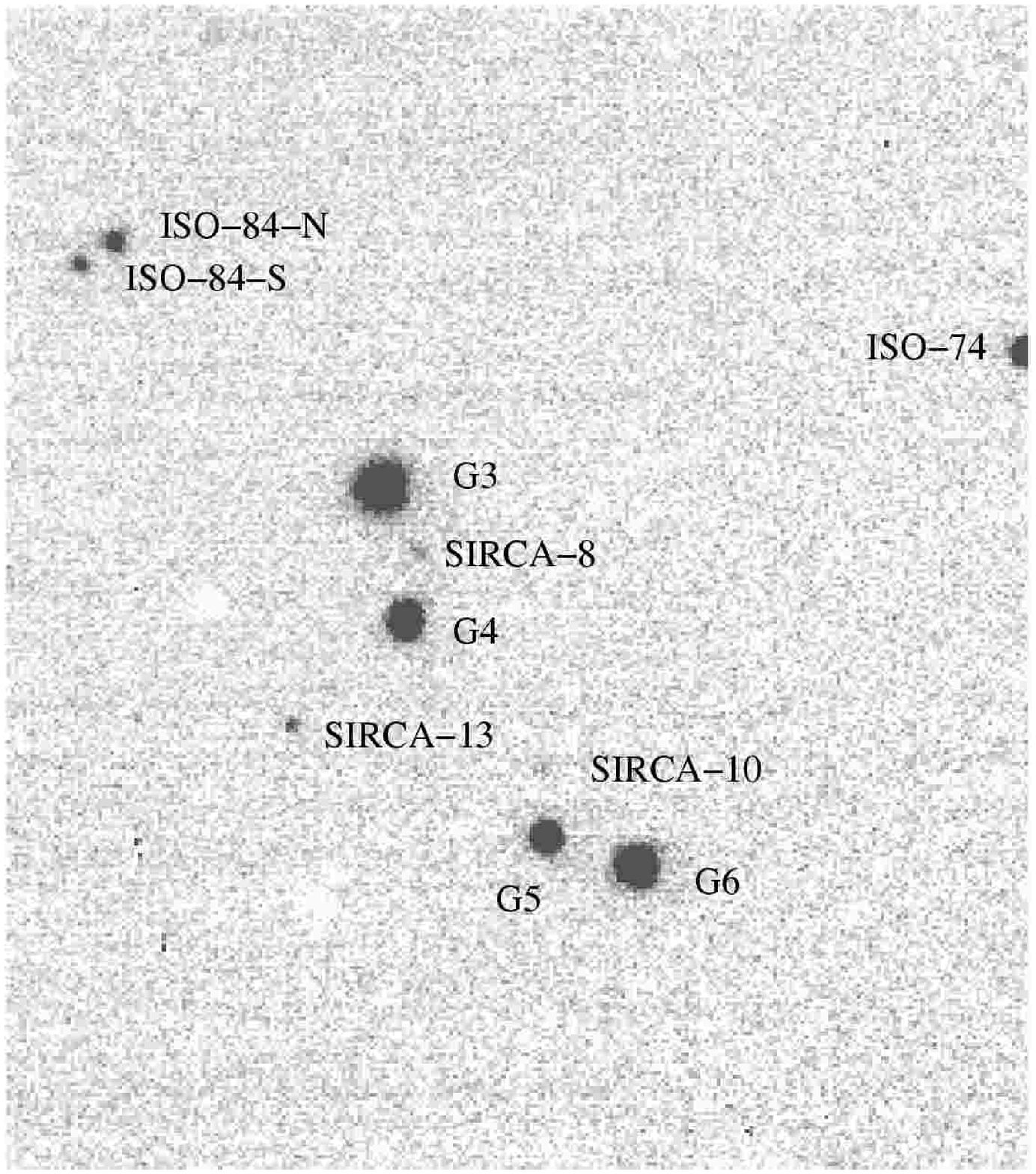}}
\caption{SIRCA/NOT $L'$ band image of a 70'' $\times$ 56'' area around 
the Ser/G3-G6 group. North is up and East to the left. }
\label{fig-g3g6im}
\end{figure}

The Stockholm Infrared Camera 
(SIRCA\footnote{See \citet{olo99} and \citet{vdb00}, as well as URL
http://www.astro.su.se/groups/infrared/instrumentation.html for more 
information on SIRCA.}) was used at the NOT, La Palma, 15-19 July 
2003. We imaged 12 SIRCA fields (56'' $\times $ 70'') within the 
ISOCAM map in Serp-NH3. 
This instrument/telescope combination has a sensitivity in the $L'$ 
band (3.8 $\mu$m) of 0.2 mJy per square arcsecond, i.e., a limiting 
magnitude of $L'$ = 15.2. The detector is a 256 $\times$ 320 InSb 
array, and the spatial resolution at the NOT is 0.22''/pix. The 
images were taken with an individual integration time of 0.2 seconds, 
5 reads per chopper position, 5 chopper repetitions, and 20 ABBA 
``noddings'' of the telescope, i.e., a total on-source integration 
time of $0.2 \times 5 \times 2 \times 5 \times 20 \times 4 = 800$ 
seconds. The bright UKIRT standards HD161903 and HD203856 with $L'$ 
magnitudes 7.01 and 6.84 \citep{cas92}, respectively, were integrated 
80 seconds (only two telescope noddings). 
The flatfield was obtained from the target frames by chopping 
between the sky and outside the field, i.e., the cold interior of 
the instrument, which effectively subtracts the dark current. Target 
images are shifted and added for each nodding cycle and flatfield 
corrected.

\begin{table}
\caption{The 46 SIRCA L' band detections in 12 pointed observations in
the Ser/G3-G6 complex. Only available in electronic form at the CDS.
\label{tab3}
        }
%
%
\end{table}

Our pointed observations with SIRCA gave $L'$ band photometry of 46
sources in 12 fields. All these are listed in Table~\ref{tab3}. The
columns give SIRCA ID number, RA and DEC (J2000) 
positions\footnote{The RA and DEC (J2000) positions of SIRCA sources
are taken to be the 2MASS position. If no 2MASS counterpart is found,
then the position is calculated with respect to other sources in the
image.}, L' magnitude and its error, K$_S$ magnitude and its error, 
2MASS ID, Arnica/NOT K$_S$ ID, ISO-NH3 ID, and other ID. The $K_S$ 
band magnitudes are taken from 2MASS except for IDs 8, 15, 16, 21, 
22, and 34, where they are taken from the Arnica/NOT observations 
listed in Table~\ref{tab2}. Positions are from 2MASS when a 2MASS 
source is available, otherwise they are ISO positions or positions 
found from the $L'$ band image with respect to other sources. For IDs 
21 and 22 the positions are from the Arnica/NOT images.

Five repeated measurements of the standard star HD161903 over 4 
nights gave a rms deviation of 0.048 magnitudes. No correction was 
made for airmass, since this would be smaller than the variable 
extinction produced by the presence of dust from Sahara at the time 
of observing. The standards were relatively near to the targets, and 
the instrumental target magnitudes have been calibrated with the 
standard observation nearest in time. Magnitudes are calculated with 
the aperture photometry task {\em phot} in IRAF, using 20 pix (4.4'') 
aperture radius for standards and 5 pix (1.1'') for program stars. 
For the small aperture an aperture correction was applied to 
compensate the loss of flux in the wings. The correction was 
determined individually in each image, apart from a few cases where 
this was not possible and the standards were used instead.
The field centred on Ser/G3-G6 is shown in Fig.~\ref{fig-g3g6im}.

\subsection{H$_2$ line imaging with NOTCam at the NOT}
\label{obsnotcam}

The Nordic Optical Telescope near-IR Camera/spectrograph (NOTCam) 
was used to do narrow-band imaging in the H$_2$ line (S(1) 1-0 at 
2.122 $\mu$m) and a nearby continuum ($\lambda_{c}$ = 2.087 and
FWHM = 0.02 $\mu$m) at the location of the NH$_3$ core to the 
northeast of the Ser/G3-G6 group. The detector was the 1024 
$\times$ 1024 $\times$ 18$\mu$m Hawaii engineering grade array. 
Ser/G3-G6NE was observed on 28 May 2003. The wide field camera
(0.235''/pix) was used, and the observations were 
performed with a ramp-sampling readout mode. We integrated for a 
total of 100s per sky position, reading the array every 10s and 
using the linear regression result of these 10 readouts to lower 
the read noise. Altogether 20 dithered sky positions were
observed, giving a total of 2000\,s on-source for each of the 
two filters. 

Flat fields were observed on the twilight sky creating
{\em bright\,-\,faint} pairs for a proper subtraction of the thermal, 
dark, and stray light contributions. 
Figure~\ref{fig-h2} shows the H$_2$ line (2.122 $\mu$m) image of 
the Ser/G3-G6NE region, and Fig.~\ref{fig-cnt} shows the narrow-band 
continuum ($\lambda_{c}$ 2.087 $\mu$m) image of the same region.
The images suffer from some stripy pattern due to pick-up
noise in the electronics at the time of observing. However, it is 
clear that most of the extended emission is pure H$_2$ line emission. 
Only around the position of ISO-94 and ISO-101 (cf. 
Fig.~\ref{fig-h2-lw2}) faint, extended continuum emission is seen.

The H$_2$ line image is flux calibrated using the near-IR standards 
AS33-0 and AS33-1 \citep{hun98}. The zeropoint difference between
the H$_2$ and the $K$ filter was found to be 2.635 mag. We transform 
to monochromatic flux through the relation F$_{\nu}$ = 710 $\times$ 
10$^{-0.4 \times {\rm m}}$ (Jy) for a zero$^{th}$ magnitude star at 
2.122 $\mu$m, assuming the stars' magnitudes are the same at 2.122 
as at 2.2 $\mu$m. From the H$_2$ line image of the standards, we get 
a conversion factor of 15.75 $\mu$Jy per ADU/s for the image. The 
intensities in the knots range from 2 to 46 $\mu$Jy/pix, and the 
noise in the background is $\sigma_{sky}$ = 0.5 $\mu$Jy/pix.

\subsection{IRAM 30m telescope observations}
\label{obs-iram}

A 1.3~mm dust continuum mosaic of part of the ISOCAM coverage was 
taken with the IRAM 30-m telescope equipped with the MPIfR 37-channel 
bolometer array MAMBO-I \citep{kre98} during four nights of observing 
sessions in March 1998. The passband of the MAMBO bolometer array has 
an equivalent width $\approx 70$~GHz and is centred at $\nu_{eff} 
\approx 240$~GHz. 

The mosaic consists of 18 individual on-the-fly maps that were 
obtained in the dual-beam raster mode with a scanning velocity of 
8$\arcsec$/sec and a spatial sampling of 4$\arcsec$ in elevation. 
In this mode, the telescope continuously scans in azimuth 
along each mapped row while the secondary mirror wobbles in 
azimuth at a frequency of 2~Hz. A wobbler throw of 45$\arcsec$ or 
60$\arcsec$ was used. The typical azimuthal size of individual maps 
was 4$\arcmin$. The size of the main beam was measured to be 
$\sim$~11$\arcsec$ (HPBW) on Uranus and other strong point-like 
sources such as quasars. The pointing of the telescope was checked 
every $\sim 1$~hr using the VLA position of the strong, compact 
Class~0 source FIRS1 in the Serpens Cloud Core (good to 
$\sim 0.1\arcsec $, \citealt{cur93}); it was found to be accurate 
to better than $\sim $~3$\arcsec$. The zenith atmospheric optical 
depth, monitored by `skydips' every $\sim 2$~hr, was between 
$\sim 0.2$ and $\sim 0.4$. Calibration was achieved through 
on-the-fly mapping and on-off observations of the primary calibrator 
Uranus \citep[e.g.,][ and references therein]{gri93}.
In addition, the Serpens secondary calibrator FIRS1, which has a 
1.3~mm peak flux density $\sim 2.4$~Jy in an 11$\arcsec$ beam, was 
observed before and after each map. The relative calibration was 
found to be good to within $\sim 10\%$ by comparing the individual 
coverages of each field, while the overall absolute calibration 
uncertainty is estimated to be $\sim 20\%$. 

The dual-beam maps were reduced and combined with the IRAM software 
for bolometer-array data (``NIC''; cf. \citealt{bro95}), which uses 
the EKH restoration algorithm \citep{eme79}.  

Figure~\ref{fig-iram} shows the IRAM data contour map with four bright 
sources labelled (MMS1, MMS2, MMS3, and MMS4), and Table~\ref{tab-iram}
lists the individual sources with fluxes and positions.

\subsection{VLA 3.6~cm observations}
\label{obs-vla}

Radio continuum observations of three fields covering all four 
millimeter sources (MMS1-4) were made at 3.6~cm (central frequency 
8.46 GHz, total bandwidth 100 MHz) with the NRAO\footnote{ The 
National Radio Astronomy Observatory is operated by Associated 
Universities, Inc., under a cooperative agreement with the National 
Science Foundation.} Very Large Array (VLA) in a mixed B/C 
configuration on 22 June 2001. The on-source integration time per 
field was $\sim 75$~min. The amplitude calibrator was 3C286 and the 
phase calibrator was 1801+010. The data were edited and calibrated 
using standard VLA procedures and the corresponding map cleaned 
using the AIPS task IMAGR with natural weighting. The synthesised 
beam size was $2.8\arcsec \times 2.3\arcsec$ (HPBW) and the primary 
beam (or field of view) of each observation was 5.3$\arcmin$ (FWHM). 
The rms noise in each field was $\sim 16 \mu$Jy/beam, implying a 
5~$\sigma$ detection threshold of $\sim 0.08$~mJy/beam in the inner 
parts of the fields. A total of eight point-like radio sources were 
detected above the 5~$\sigma$ level within the primary beams of the 
three observed VLA fields. Table~\ref{tab-vla} lists their positions 
(accurate to $\pm 0.5 \arcsec$) and their 3.6~cm flux densities 
corrected for primary beam attenuation. Five VLA sources appear 
to be closely associated with infrared and/or millimeter sources 
(cf. Table~\ref{tab-vla}). The other sources are probably background 
extragalactic sources. In addition to these robust VLA detections, 
we have also included a weak, tentative ($\sim 3.3\, \sigma$) radio 
source in Table~\ref{tab-vla} as it coincides within 1.5$\arcsec$ 
with the 1.3~mm continuum source MMS1-c.


\section{IR-excess sources}
\label{irex}

The results from the different datasets on the ISOCAM selected sample 
are summarised in Table~\ref{tab1}, which is available at the CDS in 
electronic form only. Its first column lists the ISO identification 
number (ISO-nh3-\#), then the ISOCAM positions (RA and DEC in J2000 
epoch) in columns two and three. Columns four and five give the Arnica 
K$_S$ band magnitude and error (K$_S$(96) and $\sigma_K$). Columns
six to nine list the ISOCAM fluxes with uncertainties in mJy 
($F_{\nu}(6.7 \mu)$, $\sigma_{6.7}$, $F_{\nu}(14.3 \mu)$, 
$\sigma_{14.3}$). Columns 10 to 13 show the ISOCAM detection and 
photometry flag, the name of the 2MASS counterpart, the id of the NOT 
(Arnica K$_S$(96)) counterpart, and at last a column with possible 
other identification. We refer to the 2MASS catalogue for further 
information on the 2MASS counterparts.

In the following subsections we will use various colour-magnitude 
and colour-colour diagrams to extract YSO candidates based on IR 
excess. The strategy is as follows:
\begin{itemize}
\item select IR-excess sources from the ISOCAM data 
\item compare with the $J-H/H-K_S$ diagram from 2MASS and search
      for more candidates
\item combine bands of ISOCAM and 2MASS to search for additional 
      IR-excess sources in the $H-K_S/K_S-m_7$ diagram 
\item combine SIRCA L' band with 2MASS $H$ and $K_S$ to search 
      for more candidates 
\end{itemize}

\subsection{The ISOCAM colour-magnitude diagram}
\label{irexiso}

\begin{figure}
\resizebox{\hsize}{!}{\includegraphics{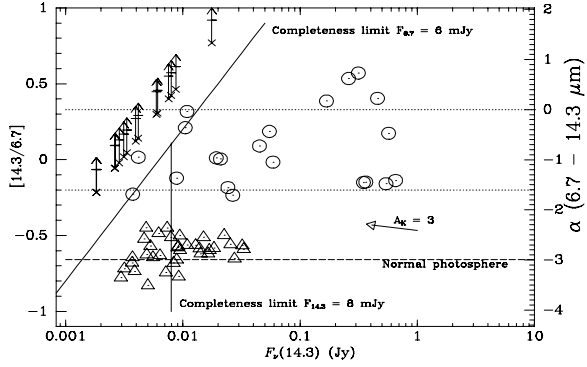}}
\caption{The colour index $[14.3/6.7]$,
defined as $\log ((F_{\nu}(14.3 \mu)/F_{\nu}(6.7 \mu))$, is shown 
on the y-axis and the flux at $14.3 \mu$m (in Jy) is shown on the 
x-axis. Sources without IR excess tend to line up around 
$[14.3/6.7] = -0.66$ or $\alpha = -3$ (dashed line). The 21 sources 
with mid-IR excesses (circles) are spread out over a large range in 
brightness. These are all YSOs. Lower limits on colour are given
for sources detected at 14.3 $\mu$m only (small crosses).}
\label{fig-323}
\end{figure}

For the ISOCAM sources we present a colour-magnitude diagram in 
Fig.~\ref{fig-323}. The colour index $[14.3/6.7]$, defined as 
$ \log (F_{\nu}(14.3 \mu m)/F_{\nu}(6.7 \mu m))$, is plotted against 
the flux in the 14.3 $\mu$m filter (magnitude indicated on upper 
x-axis). This colour index can be related to the spectral energy 
distribution (SED) index 
$\alpha = -(d \log \lambda F_{\lambda})/(d \log \lambda)$ 
calculated between 6.7 and 14.3 $\mu$m, which is indicated on the 
right-hand y-axis. As seen in all star formation regions surveyed 
with ISOCAM, the sources tend to separate into two distinct groups: 
sources for which the colour index follows that of normal stellar 
photospheres, located around $[14.3/6.7] = -0.69$ or $\alpha = -3$, 
and indicated with a dashed line in Fig.~\ref{fig-323}, and sources 
with intrinsic IR excess. Among the 57 sources with fluxes in both 
bands, 21 have mid-IR excesses. These are all YSO candidates. We 
note that they occupy a broad range in brightness (4-900 mJy at 
6.7 $\mu$m). In addition, their amount of IR excess at 14.3 $\mu$m 
is apparently independent of brightness.

There are 16 sources detected at 14.3 $\mu$m only and plotted as 
upper limits (small crosses) in Fig.~\ref{fig-323}. Only 8 of these 
have 2MASS counterparts, all with good photometric quality in $K$. 
The SED index taken between 2 and 14 $\mu$m (cf. Sect.~\ref{ysopop}) 
gives $-1.01 < \alpha_{\rm IR}^{2-14} < 1.2$, values corresponding 
to an amount of IR excess typical of Class\,I and II types of YSOs. 
These are: ISO-60, 136, 143, 153, 166, 175, 181, and 182,
of which ISO-143 is extremely red and outside of the plotted region 
in Fig.~\ref{fig-323}. We add these 8 to our list of YSO candidates. 
The remaining 8 might be spurious sources.

There are as many as 103 sources detected at 6.7 $\mu$m only. For 97 
there is a 2MASS counterpart, and all but 6 of these have good 
photometric quality in all three bands. These sources are discussed 
in Sect.~\ref{irexhk7}.

\begin{figure}
\resizebox{\hsize}{!}{\includegraphics{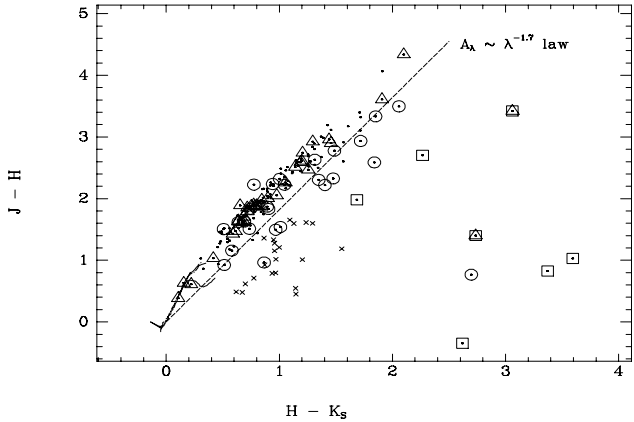}}
\caption{The $J-H/H-K_S$ diagram for the ISOCAM sources with 2MASS
counterparts (dots). The red ISOCAM sources (open circles) merge 
together with the blue ISOCAM sources (open triangles) along the 
reddening band of normal stars. Seven ISOCAM sources appear to have 
excess emission in the $H-K_S$ index (open squares), but they are 
all flagged 2MASS sources. Twenty faint 2MASS sources (crosses) 
with apparent IR excesses are discussed in the text.
}
\label{fig-jhk}
\end{figure}   

\subsection{The 2MASS $J-H/H-K_S$ diagram}
\label{irexjhk}

Figure~\ref{fig-jhk} shows the $J-H/H-K_S$ diagram for all the 
ISOCAM sources with 2MASS counterparts (dots). The blue ISOCAM
sources are marked with open triangles. The red ISOCAM sources
found to be YSO candidates in the previous section are marked 
with open circles. The 
loci of giant, supergiant, and main-sequence stars \citep{koo83} 
are indicated with bold curves. We have calculated the slope 
outlined by the ISOCAM blue sources to be 1.82$\pm$0.05. This 
should be a good indicator of the cloud extinction in this 
diagram, and the reddening vector for an A0 star is shown by 
the dashed line. For the 2MASS $JHK_S$ filter pass bands at 
$\lambda \lambda$ 1.235, 1.662, and 2.159, respectively, this 
slope is in accordance with the $A_{\lambda} \propto 
\lambda^{-1.7}$ parametrization of the NIR extinction law 
\citep{whi88}. This same opacity index $\beta = 1.7 \pm 0.4$ 
was found for Serpens in the $J$ to $H$ wavelength region by 
\citet{fro04}, also based on 2MASS data. 

\begin{figure}
\resizebox{\hsize}{!}{\includegraphics{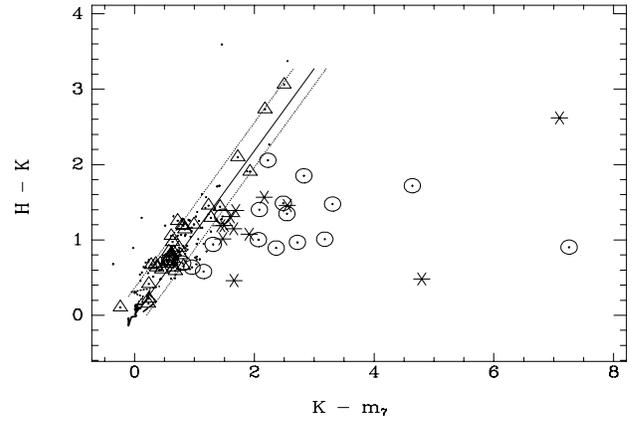}}
\caption{The $H-K_S/K_S-m_7$ diagram for 151 ISOCAM sources. The
reddening vector indicated for an A0 star (thin continuous line)
is found by a linear fit to the blue ISOCAM sources (open triangles).
The dotted lines indicate the reddening band of giants, supergiants,
and main-sequence stars. The red ISOCAM sources (open circles) are 
practically all to the right of the reddening band. The 12 
additional sources with IR excess found from this plot are marked 
as large asterisks.}
\label{fig-hk2}
\end{figure}

Figure~\ref{fig-jhk} shows that the majority of the sources are 
clustered along the reddening band. To distinguish IR excess 
sources from reddened sources in the $J-H/H-K_S$ diagram, they
must be located to the right of the reddening line by at least
1 $\sigma$ of the individual uncertainties in the $H-K_S$ index.
Only 10 of the 29 mid-IR excess sources have a detectable IR 
excess in the near-IR, as measured from the $J-H$ and $H-K_S$ 
colours. This is in line with the findings of, e.g., 
\citet{kaa00}, that only half or even less of the IR-excess 
population found from mid-IR colours is sampled by using near-IR 
colours. 

There are 7 ISOCAM sources, for which no IR excess could be 
found from the ISO observations, which seem to have excesses in 
their $H-K_S$ color (see Fig.~\ref{fig-jhk}). We note that all of 
these are upper limits in at least one of the 2MASS bands and 
should not be considered YSO candidates. Two of them are blue 
ISOCAM sources (ISO-59 and 116), one is independently found to have 
IR excess (ISO-64, see next section), but the nature 
of the remaining 4 cannot be assessed (ISO-31, 53, 91, 128). 

In the ISOCAM field there are 1221 2MASS sources with photometric 
quality flag\footnote{See the 2MASS catalogue description at
http://www.ipac.caltech.edu/2mass/releases/allsky/doc/ for details.}
A to D in all three $JHK_S$ bands. Among these there are many 
faint sources not detected by ISOCAM with apparent near-IR excesses. 
Because these are faint sources at the survey limit, we consider 
those located to the right of the reddening band by at least 2 
$\sigma$ of the individual uncertainties in the $H-K_S$ index. The 
resulting 20 sources (see Fig.~\ref{fig-jhk}) occupy a 
range of $J-H$ and $H-K_S$ colours typical of pre-main sequence 
stars. As suggested by the referee, we cross-correlated 
these with the on-line c2d Spitzer photometry table for Serpens, 
and found IRAC counterparts for 19 sources. None of these have 
IR excess in any of the Spitzer IRAC bands, however, 
and therefore they cannot be proposed as bona-fide IR excess 
YSO candidates. The apparent excesses in $H-K_S$ are not 
understood and may mostly be the result of statistical fluctuations 
in the 2MASS data, since none of the 20 sources is located to the 
right of the reddening band by more than 5 $\sigma$ of the individual 
errors in the $H-K_S$ index. More data is needed to establish
the near-IR colours of these sources.

\subsection{The $H-K_S/K_S-m_7$ diagram}
\label{irexhk7}

As seen in Fig.~\ref{fig-jhk}, the $J-H/H-K_S$ diagram does not 
extract nearly half of the IR-excess sources identified with 
ISOCAM. The IR excess does not separate well from the reddening 
in the $J-H$ and $H-K_S$ colours, and the blue and red ISOCAM 
sources tend to merge together along the reddening band. As 
shown in, e.g., \citet{kaa04}, the $H-K_S/K_S-m_7$ diagram better
distinguishes IR excess from reddening. By using this 
diagram we can study the 103 ISOCAM sources for which there
are 6.7 $\mu$m fluxes only (see Fig.~\ref{fig-hk2}). The 
reddening vector indicated is empirically found from this dataset. 
The slope 1.09 $\pm$ 0.01 is found by a linear fit to the blue 
ISOCAM sources. For comparison, the slope was 1.23 for the Serpens 
Cloud Core \citep{kaa04}. The width of the reddening band of giants, 
super giants and main-sequence stars, whose intrinsic colours are 
outlined by the bold face curves, is indicated. The red ISOCAM 
sources are all, except two (132 and 167), found to the right of 
the reddening band. 
We find a total of 26 IR-excess objects from the $H-K_S/K_S-m_7$ 
diagram, i.e., those who are located to the right of the reddening
band of normal stars by more than 1 $\sigma$ in the $K_S-m_7$
colour index. Twelve of these are new, i.e., not selected in the 
colour diagrams discussed above.

\begin{figure}
\resizebox{\hsize}{!}{\includegraphics{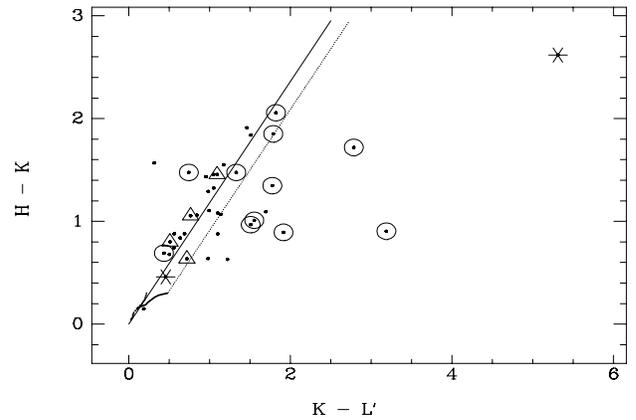}}
\caption{The $H-K_S/K_S-L'$ diagram for 39 sources (dots). We show 
by overplotting symbols the red ISOCAM sources (open circles), blue 
ISOCAM sources (open triangles), and the additional IR-excess 
sources identified in the $H-K_S/K_S-m_7$ diagram (large asterisks). 
The reddening band of normal stars is indicated (dotted line) as is 
the reddening line for an A0 star (thin continuous line). Four new 
IR excess sources are identified.}
\label{fig-hkl}
\end{figure}

\subsection{The $H-K_S/K_S-L$ diagram}
\label{irexhkl}

The $H-K_S/K_S-L'$ diagram is shown in Fig.~\ref{fig-hkl}. We plot 
the 39 sources available in these colours, and overplot red ISOCAM 
sources, blue ISOCAM sources, and additional IR-excess sources 
found from the $H-K_S/K_S-m_7$ diagram. The intrinsic colours of 
main sequence and giant stars \citep{bes88} are indicated, too. The 
reddening vector of an A0 star is drawn applying the $A_{\lambda} 
\propto \lambda^{-1.7}$ law \citep{whi88} again, giving a slope of 
1.18. The reddening band for normal stars is also indicated, and IR
excess sources must be located to the right of this line by more
than one $\sigma$ of the individual errors in the $K_S-L'$ colours.
We note that the distinction between the red and blue ISOCAM sources
is not so clear in this diagram as in the $H-K_S/K_S-m_7$ diagram, 
but significantly clearer than in the $J-H/H-K_S$ diagram. There are
no blue ISOCAM sources in the IR-excess locus, but there are five
red ISOCAM sources in the locus of reddened normal stars. In total
11 sources have IR excess in the $K_S-L'$ colour, of which four were
not previously identified as IR-excess objects. None of these four
is in the ISOCAM selected sample of Table~\ref{tab1}, but two of
them are the objects G4 and G5, which are very close to G3 (ISO-82)
and G6 (ISO-78), respectively. The third object is in the same area
just SE of G4, and the fourth object is a faint star in the SIRCA 
field of ISO-167.

\subsection{Extracted YSO candidates}
\label{irexsum}

From the 186 ISOCAM detections we have extracted {\bf 29} YSO
candidates from the ISOCAM data, and 12 additional from the 
combination of ISOCAM and 2MASS using the $H-K_S/K_S-m_7$ 
diagram. This gives a total sample of 41 IR-excess YSOs 
within the ISOCAM selected sample. In addition, we find 4 
IR-excess objects from the $H-K_S/K_S-L'$ diagram. 
Summing up, this gives a total of $45$ YSOs.

\begin{figure}
\resizebox{\hsize}{!}{\includegraphics{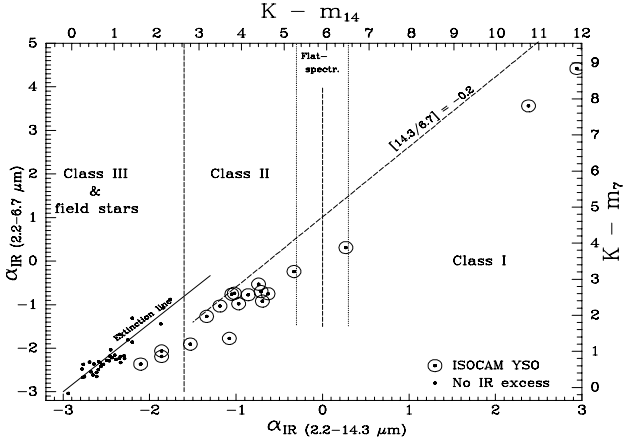}}
\caption{The spectral index $\alpha_{IR}^{2-7}$ versus 
         $\alpha_{IR}^{2-14}$ or equivalently, the $K-m_{7}/K-m_{14}$ 
         colour diagram. The locations of Class\,I, Class\,II, and 
         Class\,III sources and field stars are indicated, as well 
         as the flat-spectrum source locus.}
\label{fig-k2k3}
\end{figure}

%
\section{The YSO population}
\label{ysopop}

According to the current empirical picture of young stellar 
evolution, low mass YSOs go through evolutionary phases with 
different Spectral Energy Distribution (SED) characteristics: 
Class\,0, Class\,I, (flat-spectrum,) Class\,II, and Class\,III 
sources \citep{ada87,lad87,and93,and94}.

The Class\,0 sources are in the deeply embedded main accretion
phase and have circumstellar envelope masses larger than their
central stellar mass. The defining characteristics of a Class\,0 
is: strong, centrally condensed dust continuum emission in the 
submm, very little emission shortward of 10 $\mu$m, and powerful 
jet-like outflows \citep{and00}. Thus, we do not expect to 
sample these sources in the mid-IR, although ISOCAM did detect
a few Class\,0 sources, and Spitzer with its improved sensitivity
has detected many more. The dust continuum mapping at 1.3 mm
with IRAM should sample the Class\,0 sources, and we refer to
Sect.~\ref{flow} for a discussion of these results. The Class\,I,
flat-spectrum, and Class\,II sources, however, have IR SEDs that 
are reasonably well sampled by flux measurements at 2.2, 6.7, 
and 14.3 $\mu$m.

\subsection{Classifying the IR-excess sources}
\label{class}

We have calculated the SED index between 2.2 and 14.3 $\mu$m, 
$\alpha_{IR}^{2-14}$, which is close to the index used by 
\citet{lad84} and \citet{lad87} to define the three classes, 
I, II, and III, originally. We use the updated IR spectral 
classification scheme 
\citep{and94,gre94} and define YSOs with 
$\alpha_{IR}^{2-14} > 0.3$ as Class\,I sources, YSOs with
$\alpha_{IR}^{2-14}$ in the interval $[-0.3,0.3]$ as flat-spectrum
sources, and YSOs with $\alpha_{IR}^{2-14}$ in the interval 
$[-1.6,-0.3]$ as Class\,II sources. IR-excess sources with 
$\alpha_{IR}^{2-14} < -1.6$ are tentatively defined as Class\,III 
sources. We note that because our YSO sample is selected from IR
excesses most Class\,III objects are missed. Only a few with a
small detectable IR excess are captured, probably those objects 
in transition between clasees II and III. Class\,III sources with
no IR excess cannot be distinguished from field stars using our
data. Additional tools such as X-ray surveys must be applied to
sample the Class\,III population. 

Figure~\ref{fig-k2k3} shows the $\alpha_{IR}^{2-7}$ index versus 
the $\alpha_{IR}^{2-14}$ index with the corresponding $K-m_7$ and
$K-m_{14}$ colours in magnitudes given on the opposite axes. The 
loci of the different YSO classes are indicated.
Sources with no IR excess line up in the lower left of the
diagram along the extinction vector. The vector is taken from
\citet{kaa04} and has a length corresponding to $\sim$ 60 mag of
visual extinction. Most sources are found with $A_V$ about 10-20
mags, while one object (ISO-88 = HD170377) has no extinction and 
is perhaps a foreground star or a Class\,III cluster member at
the front side of the cloud. According to SIMBAD it is a bright
$V = 8.96$ mag optical star with spectral type K2. Sources with 
IR excess mainly occupy the Class\,II locus, while there are also 
a few in the Class\,I, flat-spectrum, and Class\,III loci. 
Compared to the Serpens Cloud Core \citep{kaa04} 
this region has fewer Class\,I and flat-spectrum objects. The 
reddest Class\,I, however, has $\alpha_{IR}^{2-14} = 2.94$, which 
is redder than any of the Class\,Is in the Serpens Cloud Core. 
Both very red objects (ISO-90 and ISO-94) are located in a core
of active star formation near the dense mm core MMS2 (see
Sects.~\ref{mms} and ~\ref{h2-line}). At least for one of these 
(ISO-94) the emission in the K-band is dominated by line emission.

We note the almost linear relationship between the two SED indices
$\alpha_{IR}^{2-7}$ and $\alpha_{IR}^{2-14}$ in Fig.~\ref{fig-k2k3}.
As in \citet{kaa04} we use this relationship to classify objects
from the $\alpha_{IR}^{2-7}$ index only, allowing us to classify
objects not detected at 14.3 $\mu$m.
From Fig.~\ref{fig-k2k3} we see that the colour $[14.3/6.7]$ is
essential to distinguish Class\,III objects with IR-excess from 
reddened stars. It is also evident that for very high extinction
there is a risk of confusing Class\,II objects with reddened stars
unless the $[14.3/6.7]$ colour is known. Nevertheless, with this
caveat in mind, we classify 12 more sources than those plotted in
Fig.~\ref{fig-k2k3}. Thus in the sample of IR-excess YSOs, we 
define as Class\,Is those that have $\alpha_{IR}^{2-7} > 1.2$ 
($K - m_7 > 4.8$), Class\,IIs those that have $\alpha_{IR}^{2-7} 
< -0.25$ ($K - m_7 < 3.2$), and flat-spectrum sources as those in 
between. 

There are, however, 6 IR-excess sources with neither of the two 
SED indices ($\alpha_{IR}^{2-14}$ and $\alpha_{IR}^{2-7}$) 
available.
In the following we will classify these sources on an ad hoc 
basis. Four of 
them are SIRCA sources (id 9, 11, 13, and 44 in Table~\ref{tab3})
with no ISOCAM detections. Plotting the various YSO classes in 
the $H-K/K-L'$ diagram we see that these four are located in the 
region occupied by the Class\,II sources in our sample. Therefore 
we tentatively classify them as Class\,IIs.
ISO-62 is a red ISOCAM source without a 2MASS counterpart located 
in the dense Ser/G3-G6SW ammonia core. Judging from the 2MASS 
sensitivity we estimate that ISO-62 has $\alpha_{IR}^{2-14} > 2$ 
and we tentatively classify it as a Class\,I object.
ISO-101, located in the dense Ser/G3-G6NE ammonia core, is a red 
ISOCAM source not detected in the near-IR continuum. 
The NOTCam images show extended emission in the H$_2$ line at 2.122 
$\mu$m and no emission in the continuum filter at 2.087 $\mu$m. 
Its rather blue $[14.3/6.7]$ colour probably also reflects a 
H$_2$ line emission (cf. Fig.~\ref{fig-h2}). Some faint extended 
emission in the L' band is barely detected at this position (see 
Table~\ref{tab3}). The lower limit estimate of $\alpha_{IR}^{2-14}$ 
first suggested that ISO-101 could be a Class\,I or 
flat-spectrum source, but considering the above discussion 
we do not count this object as a continuum source (see 
discussion in Sects.~\ref{class} and \ref{mms}).

In total, the IR-excess population we have found in the Ser/G3-G6 
complex consists of 5 Class\,I sources, 5 flat-spectrum sources, 
and 31 Class\,II sources. In addition, we see 3 Class\,IIIs with
some small IR-excesses, noting that this class is not sampled by
our IR-excess selection criterion. 

The Class\,I/Class\,II number ratio of 5/31 is more or less typical 
for star-forming regions and compares with 16/123 in $\rho$ Ophiuchi 
\citep{bon01} and 5/42 in Chamaeleon\,I \citep{kaa99b}. This is
in strong contrast to the large fraction of Class\,I sources 
found in the Serpens Cloud Core \citep{kaa04}.

\begin{table}
\caption{The 5 Class\,I and the 5 flat-spectrum sources. 
         Cross-correlated with Spitzer YSO candidates from 
         \citet{har06}. \label{tab5}
        }
\newcommand\cola {\null}
\newcommand\colb {&}
\newcommand\colc {&}
\newcommand\cold {&}
\newcommand\cole {&}
\newcommand\eol{\\}
\newcommand\extline{&&&&\eol}

\begin{tabular}{rlrrr}

\hline
{ISO} & Other ID & $\alpha_{IR}^{2-7}$ & $\alpha_{IR}^{2-14}$ & $\Delta$K \\
\extline
\hline
\cola   90\colb IRAS 18265+0028, Spitzer-16\colc  4.42\cold  2.94\cole n.a.\eol
\cola   94$^1$\colb  Spitzer-17\colc  3.56\cold  2.38\cole  0.64\eol
\cola   62$^2$\colb \colc $>$3\cold $>$2\cole n.a.\eol
\cola  143$^3$\colb \colc \cold  1.20\cole n.a.\eol
\cola   64\colb Spitzer-5\colc  2.93\cold \cole n.a.\eol
\hline
\cola   86\colb \colc  0.31\cold  0.27\cole -0.74\eol
\cola   60\colb Spitzer-3\colc \cold  0.02\cole n.a.\eol
\cola  182\colb \colc \cold -0.04\cole n.a.\eol
\cola  153\colb \colc \cold -0.14\cole n.a.\eol
\cola  121\colb \colc  1.06\cold \cole n.a.\eol
\hline

\end{tabular}

$^1$ H$_2$ line emission probably dominates the K band flux. \\
$^2$ Lower limits on $\alpha_{IR}^{2-14}$, but 
     $\alpha_{IR}^{4-14} = 1.54$ .  \\
$^3$ Not located within the 1.3 mm emission, and its Class\,I 
nature is questionable. \\

\end{table}

\subsection{Class\,I and flat-spectrum candidates}

There is a total of 10 IR-excess protostar candidates listed in
Table~\ref{tab5}: 5 Class\,Is and 5 flat-spectrum sources. We 
note that the classification is tentative in the sense that only 
lower limits on the SED indices can be given in the case of ISO-62. 
Also, the indices of ISO-94 are lower limits since the K-band flux 
is dominated by H$_2$ line emission. These sources may be even 
younger than we have indicated. Nevertheless, in the following we 
will use our tentative classification.

Two of the Class\,Is are located in the ammonia core Ser/G3-G6NE 
(ISO-90 and 94), and another two (ISO-62 and 64) in the core 
Ser/G3-G6SW. As shown in the IRAM 1.3 mm continuum map in 
Fig~\ref{fig-iram}, all but one Class\,Is are located in the high 
density regions outlined by the 1.3 mm continuum emission. The 
exception, ISO-143, lies to the NE of the two ammonia cores 
in the vicinity of ISO-141, It is extended in both ISO bands and 
not detected as a point source at 6.7 $\mu$m (therefore not 
visible in Fig.~\ref{fig-nh3-ks}). Its Class\,I nature might be
questionable. The flat-spectrum sources are partly located inside 
the cores and partly outside.

\subsection{The Class\,II and Class\,III sources}
\label{prems}

\begin{table}
\caption{The 31 Class\,II sources. The cross-correlated Spitzer 
         YSO candidates from \citet{har06} are indicated in the
         last column. \label{tab6}
        }
\newcommand\cola {\null}
\newcommand\colb {&}
\newcommand\colc {&}
\newcommand\cold {&}
\newcommand\cole {&}
\newcommand\eol{\\}
\newcommand\extline{&&&&\eol}

\begin{tabular}{rrrrl}

\hline
{ISO} & $M_J$ & $A_J$ & $\Delta K$ & Other ID \\
\extline
\hline
\cola    7\colb  4.72\colc  3.51\cold  n.a.\cole \eol
\cola   26\colb  5.67\colc  4.44\cold n.a.\cole \eol
\cola   74\colb  3.50\colc  4.88\cold 0.18\cole \eol
\cola   78\colb  3.06\colc  1.74\cold 0.05\cole CoKu-Ser-G6, NOT-158\eol
\cola   80\colb  6.06\colc  2.58\cold n.a.\cole SIRCA-5,Spitzer-10\eol
\cola   81\colb  2.80\colc  6.28\cold 0.04\cole SIRCA-7,NOT-162,Spitzer-11\eol
\cola   82\colb  2.31\colc  1.63\cold -0.11\cole CoKu Ser G3, NOT-201\eol
\cola   84$^1$\colb  5.03\colc  3.74\cold  -0.72\cole SIRCA-15 \& 16, Spitzer-14\eol
\cola   96\colb  4.76\colc  3.68\cold n.a.\cole SIRCA-23\eol
\cola  100\colb  3.90\colc  5.73\cold n.a.\cole SIRCA-28\eol
\cola  102\colb  5.67\colc  3.64\cold n.a.\cole \eol
\cola  103\colb -0.76\colc  6.70\cold n.a.\cole SIRCA-30\eol
\cola  105\colb  0.55\colc  3.47\cold 0.50\cole \eol
\cola  122\colb  3.41\colc  4.03\cold n.a.\cole \eol
\cola  136\colb  7.00\colc  4.39\cold n.a.\cole \eol
\cola  137\colb -0.28\colc  4.51\cold 0.49\cole Spitzer-21\eol
\cola  139\colb  2.42\colc  5.41\cold n.a.\cole \eol
\cola  140\colb  4.28\colc  4.49\cold n.a.\cole \eol
\cola  141\colb  4.43\colc  0.22\cold  -0.07\cole SIRCA-41,NOT-898\eol
\cola  147\colb  3.80\colc  4.55\cold n.a.\cole \eol
\cola  154\colb  5.70\colc  3.72\cold 0.25\cole Spitzer-23\eol
\cola  156\colb  4.78\colc  3.94\cold n.a.\cole \eol
\cola  166\colb  6.53\colc  3.48\cold n.a.\cole Spitzer-24\eol
\cola  175\colb  9.54\colc  0.19\cold n.a.\cole \eol
\cola  176\colb  4.46\colc  0.78\cold n.a.\cole \eol
\cola  178\colb  3.40\colc  5.18\cold n.a.\cole \eol
\cola  181\colb  6.32\colc  2.49\cold n.a.\cole \eol
\cola - \colb  4.26\colc  2.04\cold  -0.28\cole CoKu Ser G5, Spitzer-12\eol
\cola - \colb  3.23\colc  1.18\cold  -0.10\cole CoKu Ser G4\eol
\cola - \colb  7.10\colc  0.79\cold  -0.14\cole SIRCA-13\eol
\cola - \colb  5.40\colc  3.97\cold n.a.\cole SIRCA-44\eol
\hline
\end{tabular}

$^1$ Double source resolved by SIRCA and NOT/Arnica (NOT-239 \& 247), 
   but not resolved by ISO and 2MASS. \\

\end{table}

Our sample of pre-main sequence stars in the Ser/G3-G6 complex
comprises the Class\,II objects listed in Table~\ref{tab6} and
the few Class\,IIIs our selection method is capable of seeing,
listed in Table~\ref{tab7}. As emphasized in Sect.~\ref{class},
our YSO selection criterion is IR excess, and therefore we do
not sample the Class\,III population except for occasional
transition objects between Class\,II and III.

As shown in Fig.~\ref{fig-iram} the spatial distribution of the
Class\,II sources is relatively scattered, apart from the small
cluster of Class\,IIs at the location of CoKu-Ser/G3-G6.
We note that all four objects CoKu-Ser/G3-G6, earlier found 
to be T Tauri sources with strong H$\alpha$ emission \citep{coh79}
are independently classified here as Class\,II sources.
For the Class\,II sample we present $M_J$ and $A_J$ as calculated
in Sect.~\ref{lf} in Table~\ref{tab6}. The fourth column represents 
the variation in the $K_S$ magnitude between the 2MASS data 
(Jul-2000) and the Arnica/NOT (Aug-1996) dataset in the sense 
$K_S(1996) - K_S(2000)$. Among the 12 Class\,IIs where $K_S$ band 
photometry is available for both epochs, as many as 6 (or 50 \%) 
have varied by more than $\Delta K = 0.2$ mag. 

\subsection{Comparison with recent Spitzer results}

At the finalization of this paper we became aware of the recent
publication of \citet{har06} who list Spitzer mid-IR fluxes for 
24 YSO candidates found in a 12' $\times$ 12' area centred on the 
Ser/G3-G6 complex, which they refer to as Serpens Cluster B. We 
have cross-correlated these with our results and find that we 
have 4 Class\,I and 7 Class\,II sources in common, while 
\citet{har06} classify one blue ISOCAM source as a Class\,II
and find 11 YSO candidates that go undetected by ISOCAM. The 
Spitzer source number is added under {\em Other ID} in both 
Tables~\ref{tab5} and \ref{tab6}. Comparing the SED indices
with the Spitzer results, the classification is in agreement for 
all but one source, ISO-64 (Spitzer-5), where the Spitzer data
suggests it is a Class\,II. The ISOCAM 6.7 $\mu$m flux is about 
twice as high as the interpolated value between the Spitzer bands 
at 5.8 and 8.0 $\mu$m. This could be due to strong H$_2$ line 
emission in the 6.7 $\mu$m band or also to source variability.

We also note that both ISO-62 and ISO-101 are clearly detected in
the IRAC images, but not listed among the Spitzer YSO candidates 
of \citet{har06}, probably because of their extended nature.

\begin{table}
\caption{The 3 Class\,III sources with some IR-excess. Note that 
         our study does not sample the Class\,III sources. 
         \label{tab7}
        }
\newcommand\cola {\null}
\newcommand\colb {&}
\newcommand\colc {&}
\newcommand\cold {&}
\newcommand\cole {&}
\newcommand\colf {&}
\newcommand\colg {&}
\newcommand\eol{\\}
\newcommand\extline{&&&&&&\eol}

\begin{tabular}{rrrrrrr}
\hline
\noalign{\smallskip} 

{ISO} & $\alpha_{IR}^{2-7}$ & $\alpha_{IR}^{2-14}$ & {$J-H$} & {$H-K$} & {$K-L'$} & $\Delta K$ \\
\\
\hline 
\cola    2\colb -2.06\colc -1.86\cold  1.619\cole  0.638\colf  n.a.\colg n.a.\eol
\cola  132\colb -2.36\colc -2.10\cold  1.511\cole  0.735\colf  n.a.\colg n.a.\eol
\cola  167\colb -2.19\colc -1.86\cold  1.628\cole  0.691\colf  0.435\colg n.a. \eol
\hline
\end{tabular}
\end{table}

\section{Luminosity and mass function}
\label{lfmf}

\subsection{Stellar luminosities}
\label{lf}

We do not attempt to estimate luminosities and masses for the
protostar candidates (flat-spectrum and Class\,I sources). In 
this section we concentrate on the Class\,II sources.
It is probably justified to assume that our Class\,II sample 
in Table~\ref{tab6} is dominated by Classical T Tauri Stars 
(CTTS), since there are only very few examples of weak-lined 
T Tauri Stars (WTTS) with the amount of IR excess compatible 
with a Class\,II designation. 

To estimate the reddening of each object we assume an
intrinsic colour of $(J-H)_0 = 0.85$, i.e., the median value 
found for CTTS \citep{str93,mey97}. The extinction in the
$J$-band is then found from the relation $A_J = 2.53 \times 
[(J-H) - (J-H)_0]$, obtained from the $A_\lambda \propto 
\lambda^{-1.7}$ extinction law (cf. Sect.~\ref{irexjhk}). 
The absolute $J$-band magnitudes are $M_J = m_J - A_J - DM$, 
where the distance modulus is $DM = 6.76$ for the distance of 
225 pc \citep{str03}. Table~\ref{tab6} lists the Class\,II 
YSOs with extinction and absolute $J$-band magnitudes.

Apart from the photometric uncertainty in $J$, which is 
$\pm 0.05$ mag on the average, the uncertainties introduced in 
$M_J$ arise from 1) the spread in the intrinsic $(J-H)_0$ colour 
of our targets, 2) the uncertainty in the distance estimate, and 
3) the amount of veiling in the $J$-band. The spread around the 
median value of $(J-H)_0$ for CTTS is of the order of $\pm 0.15$, 
which translates directly into $\pm 0.38$ mag in the estimate of 
$A_J$. The quoted uncertainty in the distance of $\pm 55$ pc 
\citep{str03} gives an error in the distance modulus of 
$^{+0.48}_{-0.61}$ magnitudes. The average amount of $J$-band 
veiling for 22 CTTS in Taurus-Auriga was found by \citet{fol99} 
to be as high as $r = 0.57$. About 44 \% of the CTTS have this or 
a higher amount of veiling. This means that in nearly half of the 
cases, as much as 36\% or more of the flux in the $J$-band 
($r/(1+r)$) is non-photospheric. The errors introduced by not 
correcting for the veiling can thus be of the order of $0.4$ mag 
in the direction of overestimating the source brightness. On the 
assumption that these sources of errors are independent, we have 
a total uncertainty (added in quadrature) in the estimate of $M_J$ 
of $^{+0.63}_{-0.84}$ magnitudes.

\begin{figure}
\resizebox{\hsize}{!}{\includegraphics{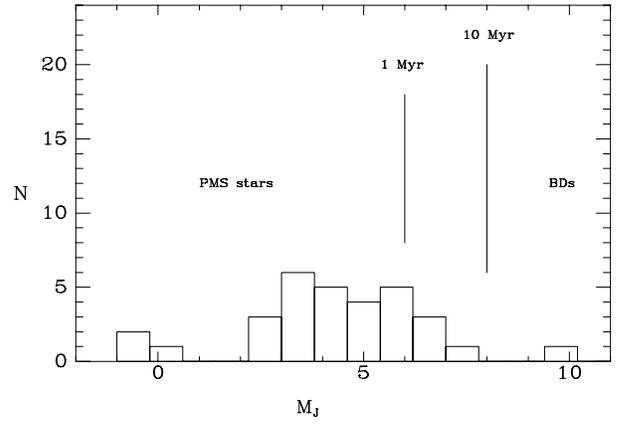}}
\caption{The $M_J$ luminosity function for the Class\,II sources 
The range of young BDs and young stars is indicated by the 
substellar limits (vertical lines) for the two ages 1 and 10 Myr.}
\label{fig-lf}
\end{figure}

The $M_J$ luminosity function of the pre-main sequence sample
is shown in Fig.~\ref{fig-lf}. A binsize of 0.8 mag was chosen 
to account for the possible uncertainties in the $M_J$ estimates 
discussed above. The vertical lines indicate the substellar limit
for the two extreme assumptions on age: 1 Myr and 10 
Myr\footnote{The age of CTTS is currently believed to be of the 
order of a few Myr, but there is some evidence of longlived 
$\sim$ 10 Myr old circumstellar disks \citep{law02,lyo03}.}.

\begin{figure}
\resizebox{\hsize}{!}{\includegraphics{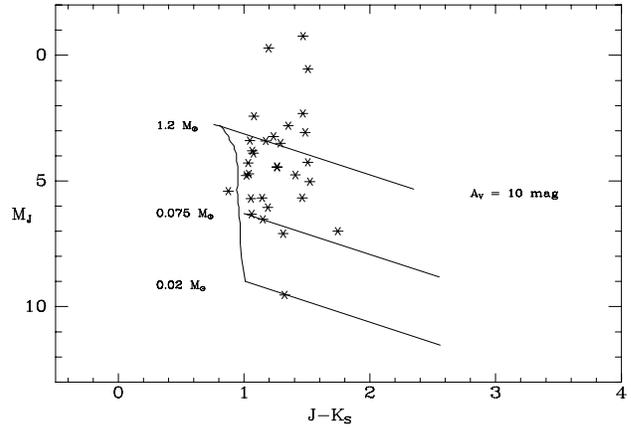}}
\caption{The $M_J$ vs. $J-K_S$ colour magnitude diagram for 
the Class\,II sources. 
The 2 Myr isochrone of \citet{bar98} and reddening vectors of
length A$_V$ = 10 mag are drawn. Note that the $J-K_S$ colour
is dereddened according to the dereddening scheme explained
in the text.}
\label{fig-jjk}
\end{figure}

The $M_J$ distribution is relatively flat over the whole 
sampled range. The completeness limit of the Class\,II sample 
is based on the completeness limit of the mid-IR survey, which 
is 6 mJy for the 6.7 $\mu$m band (cf. Sect.~\ref{obsiso}). We
have made a linear fit between the 6.7 $\mu$m flux and $M_J$ 
for the 23 Class\,II sources that have both values. The 6 mJy
completeness limit was found to correspond to $M_J = 5.6 \pm 
0.2$ mag. 

In Fig.~\ref{fig-jjk} we plot $M_J$ versus $J-K_S$ for the 31
Class\,II sources. The 2 Myr isochrone from the evolutionary 
models of \citet{bar98} is drawn with the reddening vectors 
indicated for the lowest and highest mass included in the model, 
as well as the substellar mass limit at 0.075 M$_{\odot}$. 

The $J-K_S$ colour has been corrected for reddening as discussed 
above. Because of intrinsic IR-excess and possibly also 
insufficient dereddening in some cases, practically all sources
fall to the right of the 2 Myr isochrone. 
The sample comprises objects in the mass range from above 
1.2 M$_{\odot}$ to 0.07 M$_{\odot}$ (one object reaching 
to 0.02 M$_{\odot}$), assuming an age of 2 Myr.

\subsection{Candidate young brown dwarfs with disks}
\label{bd}

The border between stars and brown dwarfs (at $\sim 0.075 
M_{\odot}$) is indicated both in Figs.~\ref{fig-lf} and 
\ref{fig-jjk}. We note that between 1 and 5 of the 31 Class\,II 
sources in Table~\ref{tab6} have substellar masses, judging 
from Fig.~\ref{fig-lf} and the age range 1-10 Myr. This same 
fraction of BDs among ISOCAM selected Class\,II objects (i.e., 
$\sim$ 20 \% for ages of a few Myr) was also found in the 
Chamaeleon I cloud \citep{olo98,per00}, the $\rho$ Ophiuchi 
cloud \citep{bon01}, and the Serpens Cloud Core \citep{kaa04}. 

All the sources in our YSO samples have IR-excess, of which 
31 sources are designated Class\,II objects. Class\,IIs have 
a mid-IR SED compatible with circumstellar dust distributed 
in a disk. 
The substellar objects in our study are therefore most probably 
young BDs with circumstellar disks. 
We note that since the selection criteria in our study is
IR excess, we do not sample any BDs without disks.

\subsection{The Serpens Class\,II mass function}
\label{mf}

The 31 Class\,II sources from this paper are added to the sample 
of 43 Class\,II sources in the Serpens Cloud Core region from 
\citet{kaa04} to improve the small number of statistics for 
the Serpens Class\,II luminosity function (LF). The two samples 
are from two different embedded young clusters separated by a 
projected distance of 45 arc minutes ($\sim$ 3 pc), but probably 
located in the same cloud system and at the same distance (225 pc
adopted from \citealt{str03}). 

\begin{figure}
\resizebox{\hsize}{!}{\includegraphics{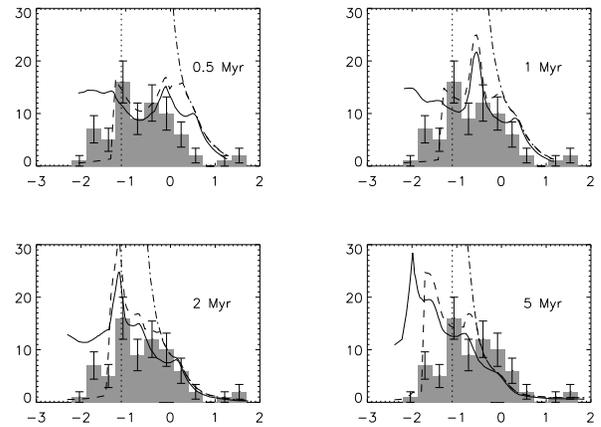}}
\caption{The histogram shows the observed LF of Class\,II
sources from the Ser/G3-G6 region (this paper) and the 
Serpens Cloud Core \citep{kaa04} as the number of sources 
per bin vs. $\log L$. Overplotted are the synthetic LFs of
a coeval population calculated on the basis of the pre-main 
sequence tracks of \citet{dan98}. Both the \citet{sca98}
three-segment power-law IMF (solid line) and the \citet{kro01}
three-segment power-law IMF (dashed line) trace well the
observed LF for a coeval population of $\sim$ 2 Myr down to
the completeness limit (vertical dotted line). The \citet{sal55} 
IMF (dashed-dotted line) has been included for reference.}
\label{fig-mf}
\end{figure}

For the 31 Class\,II sources of this paper (cf. Table~\ref{tab6}) 
we have calculated the stellar luminosity $L_{\star}$ from $M_J$
using the relation $\log L_{\star} = 1.49 - 0.466*M_J$ found by
\citet{bon01}. The total Serpens sample now consists of 74 
Class\,II sources. Figure~\ref{fig-mf} shows the observed 
luminosity function (LF) of the combined sample of Class\,II 
sources together with model LFs calculated for co-eval formation 
for a set of four ages: 0.5, 1, 2, and 5 Myr. 

We assume as a first approximation co-eval formation of the 
Class\,II sources. Though admittedly a simplification, this
assumption may be viable as we are treating a subset of
cluster members constrained evolutionarily: Class\,II 
sources are currently believed to have ages of a few Myrs. 
Also, it can be argued that star formation may proceed in several 
bursts giving rise to generations of YSOs \citep{kaa04}.

\begin{figure*}
\resizebox{\hsize}{!}{\includegraphics{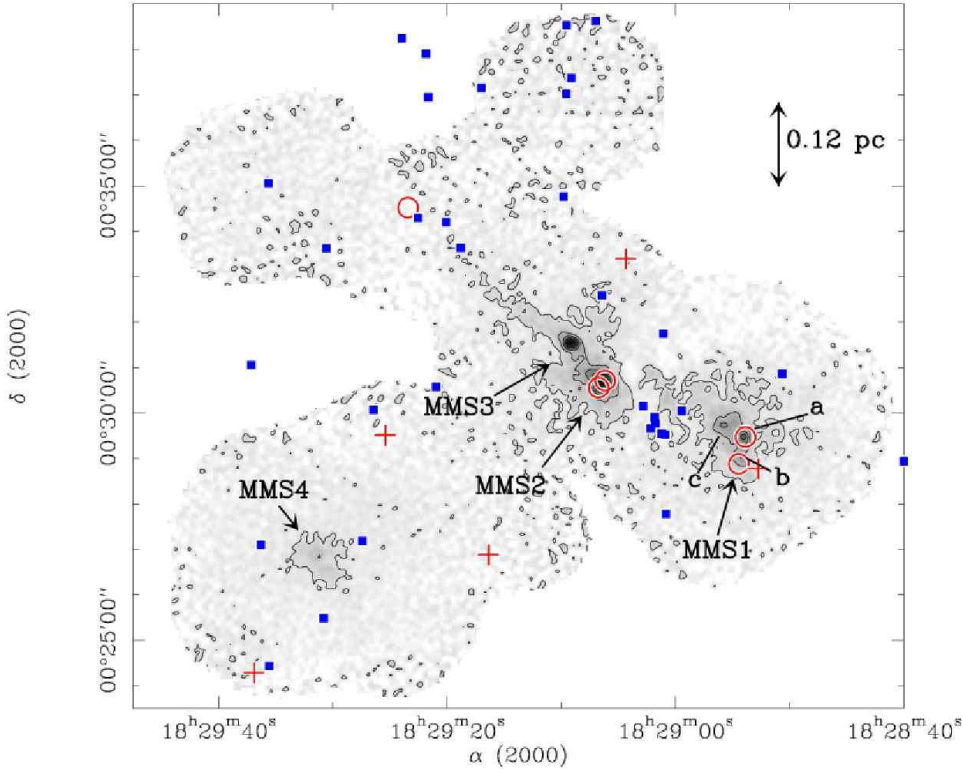}}
\caption{The IRAM 1.3 mm continuum map with the location of the YSOs 
classified in this paper marked. The contour levels go from 30 to 
450 mJy/11''-beam in steps of 60 mJy/11''-beam. The rms noise is
9 mJy/11''-beam in the best parts of the image and typically around 
15 mJy/11''-beam on average. Class\,I sources (red circles), 
flat-spectrum sources (red crosses), and Class\,II sources (blue 
filled squares) are indicated. The four bright mm sources are 
labelled MMS1, MMS2, MMS3, and MMS4. The 0.12 pc clustering scale for 
protostars found in \citet{kaa04} is shown.} 
\label{fig-iram}
\end{figure*}

For easy comparison we show the model LFs here calculated 
as in \citet{kaa04} using the pre-main sequence evolutionary 
models of \citet{dan98} and applying three different IMFs. 
Here we show the Salpeter IMF \citep{sal55}, the Scalo 
three-segment power-law IMF \citep{sca98}, and the Kroupa
three-segment power-law IMF \citep{kro01}. 
With the IMF of the form $dN \propto m^{\Gamma} d(\log m)$, 
the \citet{sal55} IMF\footnote{The
\citet{sca98} IMF has $\Gamma = -1.3$ for 
$10 M_{\odot} < m < 100 M_{\odot}$, $\Gamma = -1.7$ for 
$1 M_{\odot} < m < 10 M_{\odot}$, and $\Gamma = -0.2$ for 
$0.1 M_{\odot} < m < 1 M_{\odot}$. The \citet{kro01} IMF
has $\Gamma = -1.3$ for $m > 0.5 M_{\odot}$, $\Gamma = -0.3$ 
for $ 0.08 M_{\odot} < m < 0.5 M_{\odot}$, and $\Gamma = 0.7$ 
for $m < 0.08 M_{\odot}$, see \citet{bon06}. In \citet{kaa04} 
we instead used \citet{kro93}.} has $\Gamma = -1.35$.

In Fig.~\ref{fig-mf} the LFs produced by these IMFs for the
four ages mentioned above are shown superposed on the
observed LF, i.e., as the number of sources per bin versus
$\log L$. The completeness limit at $L \sim 0.08 L_{\odot}$
is also given (cf. Sect.~\ref{lf}). The 
normalization has been made to the number of sources above 
the completeness limit. The bin width in the histogram is 
$d \log L = 0.325$, based on a factor of two uncertainty in 
the luminosity estimate.

The increased sample in this work confirms the findings in 
\citep{kaa04} that the Serpens Class\,II LF is compatible 
with coeval formation $\sim$ 2 Myr ago and an IMF of the 
three-segment power-law type all the way down to the 
completeness limit at $L \sim 0.08 L_{\odot}$, which 
corresponds to $M \sim 0.15 M_{\odot}$ for 2 Myr.
In the plots we have extrapolated the Scalo IMF
and the Salpeter IMF to masses below which these IMFs are 
defined, i.e., beyond $0.1 M_{\odot}$ and $0.4 M_{\odot}$, 
respectively. We note that even the Salpeter IMF explains 
the data at high luminosities. 

The peak in the observed LF at $L \sim 0.09 L_{\odot}$ 
coincides with the completeness limit and cannot 
be used to constrain the age. An age dependent peak in the LF
is expected as a result of the piling up of stars in a given 
luminosity bin because of the slowing-down effect deuterium 
burning has on YSO contraction \citep{zin93}.
Nevertheless, Fig.~\ref{fig-mf} shows that neither of the three 
IMFs produce compatible LFs 
for ages much younger or older than 2 Myr. The model LFs for 
coeval populations that formed 0.5 Myr and 1 Myr ago have 
more sources than observed in the luminosity bins around 
$3 L_{\odot}$ and $0.3 L_{\odot}$, respectively. For ages 
above 2 Myr the model LFs produce fewer sources than observed 
for luminosities above $0.7 L_{\odot}$. This tendency starts 
already at an age of 3 Myr, but for 5 Myr the deviation is 
larger than the error bars for several consecutive bins in the 
observed LF. 

We thus conclude that an age of 2 Myr is a good fit for the
Class\,II population, although co-eval star formation in a
strict sense is unlikely. We cannot exclude models where 
star formation proceeds in a more continuous way. However,
we find that continuous SF over intervals shorter than 2 Myr 
is incompatible with the observed LF. A single burst of star 
formation that took place 2 Myr ago can fully explain the 
data. Bursts of short duration produce model LFs that 
resemble the co-eval LFs, while increasing the burst duration 
smooths out the structure in the LFs. 

To distinguish between the \citet{sca98} and the \citet{kro01} 
IMFs for the Serpens Class\,II sources we need to extend the 
completeness limit from $0.08 L_{\odot}$ to $0.04 L_{\odot}$. 
This should be feasible with the deeper mid-IR surveys from 
Spitzer. 
For the more nearby $\rho $ Ophiuchi region (d\,=\,140~pc), 
\citet{bon01} found the IMF of Class\,II sources (complete to 
$0.03 L_{\odot}$) to be well fitted by a two-step power-law 
with indices $\Gamma = -1.7$ for $m > 0.55 M_{\odot}$ and 
$\Gamma = -0.15$ for the whole interval $0.055 M_{\odot} < m 
< 0.55 M_{\odot}$ (accounting for unresolved binaries gives 
$\Gamma = -0.35$). Thus, there are indications that the 
Class\,II IMFs are better described by the \citet{sca98} or 
the \citet{kro93} than by the more recent \citet{kro01}.

%

\begin{figure}[t]
\resizebox{\hsize}{!}{\includegraphics{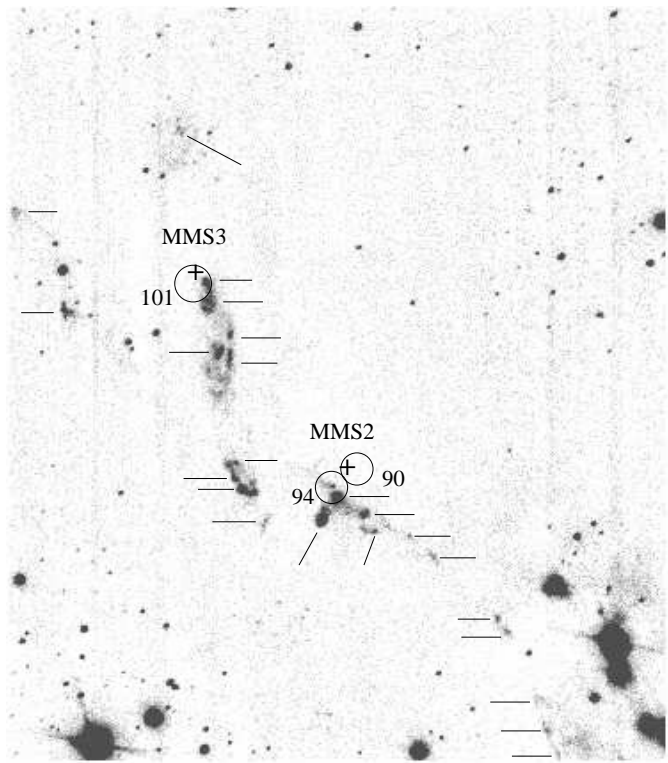}}
\caption{NOTCam $H_2$ line image (2.122 $\mu$m) of Ser/G3-G6NE. 
Pure line emission features are arrowed (cf. the continuum image
in Fig.~\ref{fig-cnt}). The locations of the two bright mm 
sources MMS2 and MMS3 are marked with plus signs. The Class\,I 
candidates ISO-90 and 94, as well as ISO-101, probably a hot
spot in the jetlike emission, are encircled. }
\label{fig-h2}
\end{figure}

\begin{figure}[t]
\resizebox{\hsize}{!}{\includegraphics{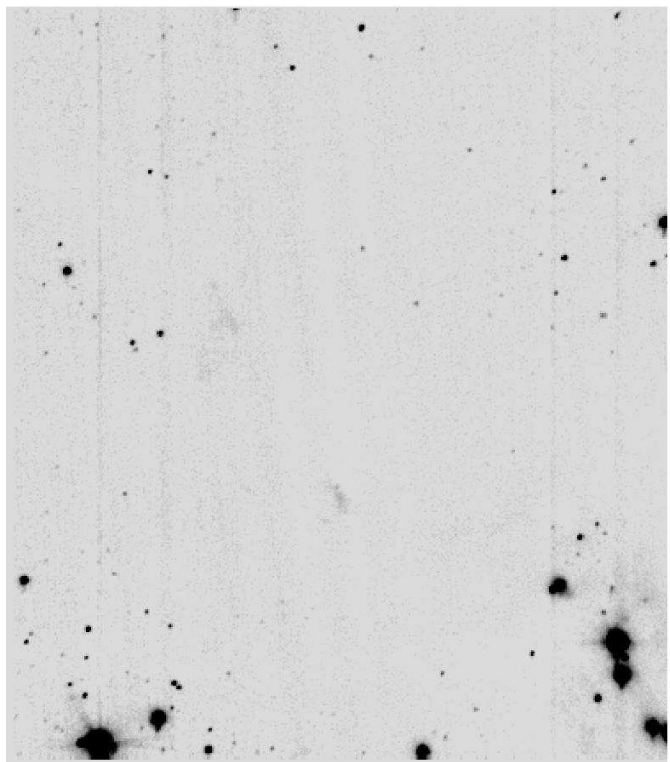}}
\caption{NOTCam 2.087 $\mu$m continuum image of Ser/G3-G6NE. The
same field as Fig.~\ref{fig-h2}.}
\label{fig-cnt}
\end{figure}

\section{Cloud structure deduced from 1.3~mm and 3.6~cm observations}
\label{flow}

Figure~\ref{fig-iram} shows the IRAM 1.3 mm continuum mapping of a 
major part of the ISOCAM field. Four strong continuum sources are
labelled in the figure. One is a multiple of three and another is 
elongated - all listed with coordinates and fluxes in 
Table~\ref{tab-iram}. The protostar clustering scale of 0.12~pc 
found in the Serpens Cloud Core \citep{kaa04} is drawn in the map 
for size reference. Also shown are the positions of YSO candidates 
found in Sect.~\ref{ysopop}: Class\,I sources, flat-spectrum sources, 
and Class\,II sources.

The contour plot of the 1.3 mm continuum emission outlines the 
position of the two ammonia cores found by \citet{cla91} on each 
side of the optically visible stellar group Ser/G3-G6 quite well.
MMS4 is a starless core coinciding with an absorption core seen 
in both ISOCAM filters. It is a candidate prestellar core similar
to the cores studied in absorption with ISOCAM by \citet{bac00}.  
MMS1 is located in the Ser/G3-G6SW ammonia core and is a multiple 
source whose brightest components we have denoted as a, b, and c. 
Both MMS2, which is extended, and MMS3 are located in the 
Ser/G3-G6NE ammonia core. 

Morphologically, the 1.3 mm map is quite similar to the IRAM map of 
the Serpens Cloud Core \citep{kaa04}, showing two separate clumps of 
which one has an elongated tail in the direction of the alignment of 
the clumps along an elongated density structure or a cloud filament. 
Here the elongated density structure oriented NE-SW is also well 
traced in the $K$-band image of Fig.~\ref{fig-nh3-ks} as a pronounced
drop in the surface density of faint stars. In the case of the Serpens 
Cloud Core, the NW-SE oriented ridge was described in, e.g., 
\citet{kaa99a}. Each clump contains a sub-structure. This region 
differs from the Cloud Core, however, in that the number of protostars 
is lower, and the 1.3 mm emission is in general less intense. The 
immediate interpretation is that we see no such intense burst of 
star formation here as in the Cloud Core, which is also in agreement 
with the Class\,I/Class\,II number fraction discussed at the end of 
Sect.~\ref{class}.

\subsection{Individual 1.3~mm and 3.6~cm sources}
\label{mms}

The fluxes and positions of the IRAM 1.3~mm and VLA 3.6~cm sources 
are given in Tables~\ref{tab-iram}~and~\ref{tab-vla}. Taking into 
account the uncertainties in the ISOCAM positions (estimated in 
Sect.~\ref{obs2m}) and the IRAM positions ($\sim 3\arcsec$ on the 
average and 5$\arcsec$ at most), here we compare the mm and cm 
source positions with those of ISOCAM. 

ISO-62 is probably the IR counterpart of MMS1-a, as the coordinates 
agree to 1.8$\arcsec$. There is no $K$-band detection, but $L' =  
11.4$ mag, which gives a mid-IR SED index $\alpha_{IR}^{4-14} = 
1.54$ -- consistent with a Class\,I source classification. The radio 
source VLA-\#2 is positioned only 3.8$\arcsec$ away from the mm 
source. The MMS1-b position differs by 8.5$\arcsec$ from the IR 
coordinates of ISO-64, which is also a Class\,I source. Since this 
is just below the 3$\sigma$ in positional uncertainty, it can not 
be excluded that ISO-64 is the infrared counterpart of MMS1-b. At 
the position of MMS1-c we find no IR counterpart, although we note 
that this region suffered strongly from persistency effects in both 
ISOCAM maps. There is, however, a 3.3 $\sigma$ radio source 
(VLA-\#3) only 1.4$\arcsec$ to the south of MMS1-c. We tentatively 
suggest MMS1-c as a possible Class\,0 candidate. 

MMS2 is elongated, and its position determined by the peak emission 
is found between the ISOCAM positions of the two Class\,I sources 
ISO-90 and ISO-94. The MMS2 elongation is roughly in the direction 
towards the mid-IR sources, and we interpret these to be the IR 
counterparts of the elongated MMS2 source, possibly an unresolved 
double. (The offsets between the peak MMS2 coordinate and the ISOCAM 
positions are 7.8$\arcsec$ and 12$\arcsec$ from ISO-90 and ISO-94, 
respectively.)  We also note that a radio source (VLA-\#6) is found 
at the position of ISO-94 offset by only 6.2$\arcsec$. We note that 
the Arnica/NOT position of the $K_S = 15.5$ mag source (NOT-343) 
agrees within 0.9$\arcsec$ with the VLA position. 

\begin{table}
\caption{The bright IRAM 1.3 mm sources with J2000 positions good 
to about $\pm 3$ arcsec. 
\label{tab-iram}
        }
\newcommand\cola {\null}
\newcommand\colb {&}
\newcommand\colc {&}
\newcommand\cold {&}
\newcommand\cole {&}
\newcommand\eol{\\}
\newcommand\extline{&&&&\eol}

\begin{tabular}{lccrr}

\hline
{Name} & $\alpha_{2000}$ & $\delta_{2000}$ & Peak S$_{1.3 \rm mm}$ & Int$^1$ S$_{1.3 \rm mm}$ \\
       &                 &                 & (mJy/beam)            & (mJy)                          \\ 
\hline
\cola  MMS1-a\colb 18 28 54.1 \colc 00 29 29\cold 234. \cole 830 \eol
\cola  MMS1-b\colb 18 28 54.3 \colc 00 29 02\cold 153. \cole 700 \eol
\cola  MMS1-c\colb 18 28 55.8 \colc 00 29 45\cold 161. \cole 780 \eol
\cola  MMS2$^2$\colb 18 29 06.5 \colc 00 30 41\cold 404. \cole 1650 \eol
\cola  MMS3\colb 18 29 09.2 \colc 00 31 32\cold 399. \cole 1020 \eol
\cola  MMS4\colb 18 29 31.2 \colc 00 26 50\cold  94. \cole 420 \eol
\hline
\end{tabular}

$^1$ Integrated over an aperture with 37$\arcsec$ diameter. \\
$^2$ This source is elongated. \\
\end{table}

\begin{figure}
\resizebox{\hsize}{!}{\includegraphics{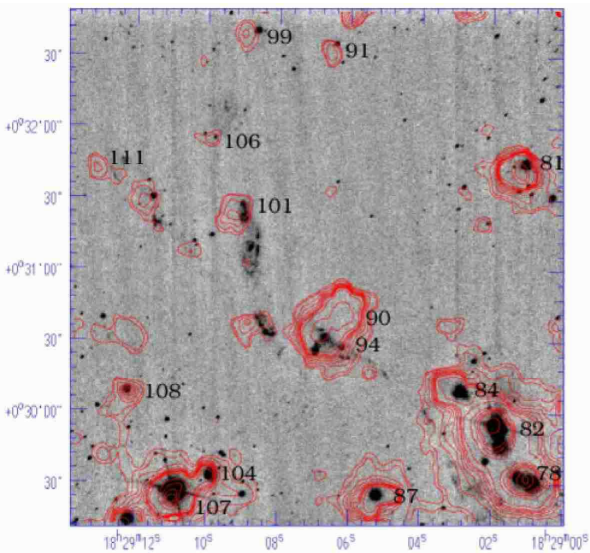}}
\caption{NOTCam $H_2$ line image (2.122 $\mu$m) of the Ser/G3-G6NE 
core. ISOCAM 6.7 $\mu$m contours are overlaid, and ISOCAM sources 
are labelled with the ISO ID number of Table~\ref{tab1}.} 
\label{fig-h2-lw2}
\end{figure}

MMS3 is separated from the faint mid-IR source ISO-101 by 6$\arcsec$. 
ISO-101 has no $K$-band detection, and in the $L'$-band only a very 
faint extended signal was seen. Judging from Fig.~\ref{fig-h2-lw2} 
the position of ISO-101 seems to coincide with the extended H$_2$ 
line emission. Also, the ISOCAM colour $[14.3/6.7]$ is rather blue 
compared to the lower limit on the $\alpha_{IR}^{2-14}$ index. This 
is likely explained by H$_2$ line emission in the 6.7 $\mu$m band 
and suggests that ISO-101 is a just a hot spot in the bright 
jet-like feature seen in the deep $H_2$ line image in 
Fig.~\ref{fig-h2}, rather than being a continuum source. Most 
probably, ISO-101 is not the mid-IR counterpart of MMS3. There is 
a radio source (VLA-\#7) less than 3$\arcsec$ from the IRAM position 
of MMS3. Based on this we propose that MMS3 is a new Class\,0 
candidate and probably the driving source of the bright outflow seen 
in the deep $H_2$ line image in Fig.~\ref{fig-h2} (see discussion 
in Sect.~\ref{h2-line}).

MMS4 has no IR sources in its vicinity, and as described above it 
is a candidate prestellar core. The VLA source \#4 is a very bright 
radio source coinciding in position with ISO-82 (CoKu Ser-G3), a 
Class\,II source that is seen optically and was classified as a 
Classical TTauri star by \citet{coh79}. The position of another 
radio source (VLA-\#8) is only 4.2$\arcsec$ away from ISO-107, a 
bright ($K = 7.8$ mag) star in the central part of the cluster 
that has no infrared excess, but could possibly be a YSO of type 
Class\,III. Three VLA sources (\# 1, 5, and 9) are without mm or IR 
counterparts and are probably extragalactic sources.

\begin{table}
\caption{The VLA 3.6~cm sources with J2000 positions good to 
$\pm 0.5$ arcsec. All except one are detections above 5$\sigma$.
The beam size is given in Sect.~\ref{obs-vla}.
\label{tab-vla}
        }

\newcommand\cola {\null}
\newcommand\colb {&}
\newcommand\colc {&}
\newcommand\cold {&}
\newcommand\cole {&}
\newcommand\eol{\\}
\newcommand\extline{&&&&\eol}

\begin{tabular}{lccrr}

\hline
{VLA} & $\alpha_{2000}$ & $\delta_{2000}$ & S$_{3.6 \rm cm}$ & Associations \\
 (\#)  &                 &                 &  (mJy/beam)&              \\ 
\hline
\cola  1\colb 18 28 53.49 \colc 00 29 11.9\cold 0.08 \cole  \eol
\cola  2\colb 18 28 54.05 \colc 00 29 32.7\cold 0.08 \cole MMS1-a, ISO-62 \eol
\cola  (3)$^{1}$\colb 18 28 55.80 \colc 00 29 43.6\cold 0.05 \cole MMS1-c \eol
\cola  4\colb 18 29 01.80 \colc 00 29 54.4\cold 3.30 \cole ISO-82, Ser-G3 \eol
\cola  5\colb 18 29 03.31 \colc 00 30 57.9\cold 0.09 \cole  \eol
\cola  6\colb 18 29 06.70 \colc 00 30 33.6\cold 0.32 \cole ISO-94, MMS2 \eol
\cola  7\colb 18 29 09.03 \colc 00 31 30.6\cold 0.08 \cole MMS3 \eol
\cola  8\colb 18 29 11.14 \colc 00 29 24.8\cold 0.12 \cole  ISO-107\eol
\cola  9\colb 18 29 25.52 \colc 00 25 04.0\cold 0.14 \cole  \eol
\hline
\end{tabular}

$^1$ This is a 3.3$\sigma$ detection only, and therefore only tentative.\\

\end{table}

Radio continuum emission at 3.6~cm has been detected for a number 
of YSOs -- most of them in the Class\,0 or Class\,I protostellar 
phase \citep{eir05}. Spectral indices obtained at cm wavelengths 
show that thermal free-free emission from ionised gas usually 
dominates, and it has been suggested that the radio emission from 
protostars is due to thermal radio jets ionized by shocks in strong 
stellar winds \citep{ang98,rei04}. Although we have no
information on the spectral index in the radio, this is probably 
the most likely interpretation for our VLA sources \#2, 3, 6, and 
7, which all correspond to Class\,I or Class\,0 sources, and of 
which two are associated with H$_2$ outflows (see next section). 
For comparison, in the Serpens Cloud Core \citet{eir05} found that 
as many as 10 of 16 YSOs with radio emission were in the Class\,0 
to Class\,I phase, while only about 25\% were as evolved as 
Class\,II sources. For this last group it has been suggested that
non-thermal gyrosynchrotron emission from coronally active stars 
may explain the radio emission \citep{smi99}. We speculate that 
this could be the case for VLA~\# 4 and 8, although we have no 
X-ray information for these sources. Determination of the radio 
spectral indices is needed to confirm the above suggestions.

\subsection{Outflows traced by the 2.122 $\mu$m S(1) line of H$_2$}
\label{h2-line}

\begin{figure}
\resizebox{\hsize}{!}{\includegraphics{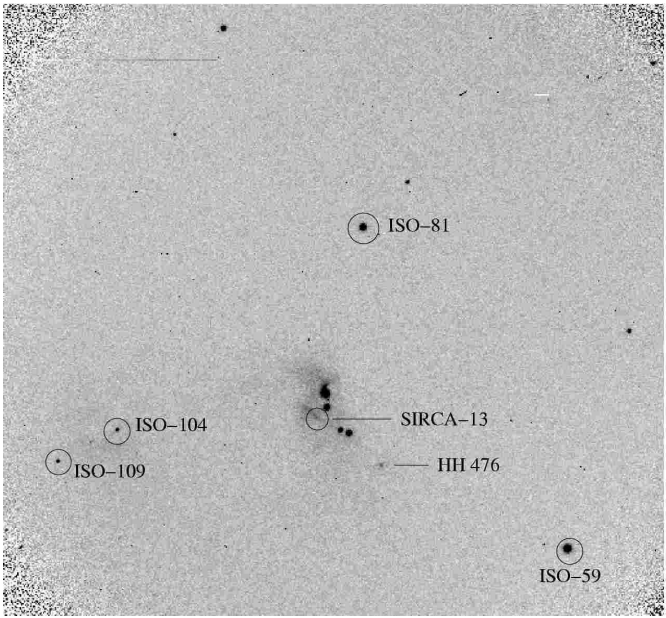}}
\caption{ALFOSC/NOT image of 900s integration through the 
H$_{\alpha}$ filter (18 nm wide) in June 2002. Apart from some 
extended emission around the Ser/G3-G6 group, the only pure line 
emission object seen in the optical is the already known HH 476. 
FOV=6.5'. }
\label{fig-ha}
\end{figure}

The Ser/G3-G6NE ammonia core was imaged with NOTCam in the 2.122 
$\mu$m S(1) line of H$_2$ (see Fig.~\ref{fig-h2}). Numerous 
Herbig-Haro (HH) objects with pure H$_2$ line emission are found 
by comparison with the 2.087 $\mu$m continuum image in 
Fig.~\ref{fig-cnt}. The morphology is complex, and proper motion 
of the knots is not yet available, but the features probably arise 
from at least two different bipolar collimated flows crossing 
each other along the line of sight. From the optical H$_{\alpha}$ 
image in Fig.~\ref{fig-ha}, it is evident that the flows are deeply 
embedded. We tentatively suggest that the driving sources 
are the two dense mm cores MMS2 and MMS3, the last one proposed to 
be a new Class\,0 candidate.

Among the two bipolar flows traced by the numerous HH objects seen 
in the near-IR in Fig.~\ref{fig-h2}, one is a very long collimated 
flow oriented NE-SW and extending over the full FOV of the image. 
It probably arises from MMS2, either of the possible multiple or 
as a consequence of disintegrating multiples, a mechanism 
suggested by \citet{rei00} and \citet{rei04}. Its extension is 
at least 90$\arcsec$ to both sides of the driving source. We 
notice that the flow is slightly S-shaped. In the NE it ends in 
a bow-shock-like morphology (upper left of image). The optical 
Herbig-Haro object HH~476 that falls just outside the H$_2$ 
image (but is shown in the larger field H$_{\alpha}$ image in 
Fig.~\ref{fig-ha}) may be related to this flow. HH 476 was 
detected by \citet{zie99} in an optical $[S II]$ survey, and 
\citet{wu02} suggested its energy source to be IRAS18265+0028, 
which ISOCAM resolved in the two very red sources ISO-90 and 
ISO-94. Follow-up imaging is needed to determine the proper 
motion of the knots and to confirm our tentative interpretation.

The remaining H$_2$ emission features, i.e., the bright knots 
to the south of MMS3 and grossly aligned almost N-S, and also 
the faint extended emission seen to the north of MMS3 around 
the position of ISO-106, most likely arise from another 
outflow, one that is probably driven by MMS3, a new Class\,0 
candidate. The morphology is more complex with ringlike shapes, 
and more data is needed to interpret these structures. 

The only extended {\em continuum} radiation seen in 
Fig.~\ref{fig-cnt} is some faint nebulous emission in the 
vicinity of the two suggested driving sources, MMS2 and MMS3, 
and this is probably scattered light.

%
\section{Summary and conclusions}
\label{con}

An embedded cluster of YSOs was found from an ISOCAM survey around 
the optical stellar group Ser/G3-G6. Mid-IR photometry in two 
broadband filters centred on 6.7 and 14.3 $\mu$m obtained with 
ISOCAM was combined with data from various telescope/instrument 
configurations to present a multi-wavelength study. This reveals 
low-mass YSOs in the evolutionary stages from Class\,0 to 
Class\,III. We propose 2 Class\,0 candidates, 5 Class\,I candidates, 
5 flat-spectrum sources, 31 Class\,II objects, and 3 Class\,IIIs 
(with some IR-excess). Our selection criterium is IR excess, which 
means that our sample is severly incomplete for Class\,III sources in 
general. This also means that we cannot estimate the disk fraction
in the YSO population. The number fraction of Class\,I to Class\,II 
sources in the Ser/G3-G6 IR cluster (5/31) is quite typical for 
star-forming regions and comparable to the $\rho$ Ophiuchi and 
Chamaeleon~I clusters, i.e., quite different from the unusually 
large fraction of Class\,Is found in the Serpens Cloud Core 
\citep{kaa04}.

The IRAM 1.3~mm continuum mapping detects several bright continuum 
sources in the two ammonia cores Ser/G3-G6NE and Ser/G3-G6SW, and 
one outside that is suggested as a candidate prestellar core. 
MMS1-c and MMS3 are proposed to be Class\,0 candidates. Both of them
are also VLA 3.6~cm radio continuum sources. Deep NOTCam imaging in 
the 2.122 $\mu$m line of H$_2$ reveals signs of two complex outflows 
in the Ser/G3-G6NE core. A deep H$_{\alpha}$ image shows that the 
flows are 
embedded and only the previously known HH 476 is seen optically. We 
suggest that the driving sources of these two outflows are MMS2 and 
MMS3.

Comparing with existing pre-main sequence evolutionary models, we 
find that for any reasonable assumptions on age for the Class\,II 
sources, our sample extends well into the BD mass regime. The 
Class\,II sample was joined with the Class\,II sample from the 
Serpens Cloud Core to improve statistics. Model LFs calculated for 
a set of ages and IMFs give a best age of 2~Myr for which both 
the \citet{sca98} and \citet{kro01} IMFs are compatible with the 
observed LF down to the completeness limit at $0.08~L_{\odot}$. 
The Class\,II LF can be described by co-eval star formation 
$\sim$~2 Myr ago, continuous star formation over the last 2~Myr, 
or a burst of star formation of various duration that took place 
2~Myr ago. There is a peak in the observed LF that cannot be used to 
constrain the age because it coincides with the completeness 
limit. Nevertheless, the general shape of the LF does not permit 
a much younger or older population. The completeness limit must 
be shifted to $0.04~L_{\odot}$ to possibly distinguish between 
different IMFs and star formation scenarios. This should be 
feasible with the deeper mid-IR surveys now available from Spitzer.

\begin{acknowledgements}

We thank the referee, Dr Paul Harvey, for suggestions that led 
to substantial improvements of the paper. We acknowledge Dr. 
F.O. Clark for presenting his unpublished NH$_3$ mapping to us
before planning the ISOCAM observations.
The ISOCAM data presented in this paper was reduced using "CIA", 
a joint development by the ESA Astrophysics Division and the 
ISOCAM Consortium led by the ISOCAM PI, C. Cesarsky, Direction des 
Sciences de la Matiere, C.E.A., France.
This work is partly based on observations made with the NOT, 
operated on the island of La Palma jointly by 
Denmark, Finland, Iceland, Norway, and Sweden, in the Spanish 
Observatorio del Roque de los Muchachos of the Instituto de 
Astrofisica de Canarias. 
One image presented here has been taken using ALFOSC, which is 
owned by the Instituto de Astrofisica de Andalucia (IAA) and 
operated at the NOT under an agreement between 
IAA and the NBIfAFG of the Astronomical Observatory of 
Copenhagen. This publication makes use of data products from the 
Two Micron All Sky Survey, which is a joint project of the 
University of Massachusetts and the Infrared Processing and 
Analysis Center/California Institute of Technology, funded by 
the National Aeronautics and Space Administration and the 
National Science Foundation. This research has made use of the 
SIMBAD database, operated at CDS, Strasbourg, France, as well as 
SAOImage DS9, developed by the Smithsonian Astrophysical Observatory.

\end{acknowledgements}

\bibliographystyle{aa}

\end{document}